\documentclass{tlp}

\usepackage{amsmath}
\usepackage{amssymb}
\usepackage{stmaryrd}
\usepackage{pst-tree}
\usepackage[matrix,arrow,ps,dvips]{xy}
\usepackage{harvard}

\providecommand*{\Nset}{\mathbb{N}}            

\providecommand{\etal}{\textit{et al.}}

\newcommand*{\cT}{\mathcal{T}}

\newcommand*{\fund}[3]{\mathord{#1}\colon#2\rightarrow#3}

\newcommand{\defrel}[1]{\mathrel{\buildrel \mathrm{def} \over {#1}}}
\newcommand{\defeq}{\defrel{=}}
\newcommand{\defiff}{\defrel{\Longleftrightarrow}}

\newcommand{\summary}[1]{\textrm{\textbf{\textup{#1}}}} 

\newcommand{\card}{\mathop{\#}\nolimits}
\newcommand*{\sseq}{\subseteq}
\newcommand*{\sslt}{\subset}
\newcommand*{\Sseq}{\supseteq}
\newcommand*{\Ssgt}{\supset}

\newcommand*{\union}{\cup}
\newcommand*{\inters}{\cap}
\newcommand*{\setdiff}{\setminus}
\newcommand*{\bigunion}{\bigcup}
\newcommand*{\biginters}{\bigcap}

\newcommand*{\sqsseq}{\sqsubseteq}

\newcommand*{\Vars}{\mathord{\mathit{Vars}}}
\newcommand*{\vars}{\mathop{\mathit{vars}}\nolimits}

\newcommand*{\Terms}{\cT_{\Vars}}

\newcommand{\glb}{\mathop{\mathrm{glb}}\nolimits}
\newcommand{\lub}{\mathop{\mathrm{lub}}\nolimits}
\newcommand{\uco}{\mathop{\mathrm{uco}}\nolimits}

\newcommand{\st}{\mathrel{.}}
\newcommand{\itc}{\mathrel{:}}

\newcommand*{\VI}{\mathit{VI}}
\newcommand*{\SG}{\mathit{SG}}
\newcommand*{\SH}{\mathit{SH}}
\newcommand*{\sh}{\mathit{sh}}
\newcommand*{\SSl}{\mathit{SS}}

\newcommand*{\SSm}{{\mathit{SS}^\rho}}
\newcommand*{\SHm}{{\mathit{SH}^\rho}}

\newcommand*{\pairs}{\mathop{\mathrm{pairs}}\nolimits}

\newcommand*{\PS}{\mathit{PS}}

\newcommand*{\Subst}{\mathit{Subst}}

\newcommand*{\rel}{\mathop{\mathrm{rel}}\nolimits}

\newcommand*{\bin}{\mathop{\mathrm{bin}}\nolimits}
\newcommand*{\proj}{\mathop{\mathrm{proj}}\nolimits}

\newcommand*{\mgu}{\mathop{\mathrm{mgu}}\nolimits}
\newcommand*{\amgu}{\mathop{\mathrm{amgu}}\nolimits}

\newcommand*{\Defminus}{{\Def^{\mathord{-}}}}
\newcommand{\ucomeet}{\mathbin{\sqcap}}
\newcommand{\bigucomeet}{\mathop{\bigsqcap}\limits}

\newcommand*{\tuples}{\mathop{\mathrm{tuples}}\nolimits}
\newcommand*{\PSD}{\mathit{PSD}}
\newcommand*{\TSD}{\mathit{TSD}}
\newcommand*{\TS}{\mathit{TS}}

\newcommand*{\Moore}{\mathop{\mathrm{Moore}}\nolimits}
\newcommand*{\dAtoms}{\mathop{\mathrm{dAtoms}}\nolimits}
\newcommand*{\MI}{\mathop{\mathrm{MI}}\nolimits}
\newcommand*{\idc}{\mathit{id}_{\scriptscriptstyle C}}

\newcommand*{\PSDplus}{{\PSD^+}}
\newcommand*{\PSDplusplus}{{\PSD^{\mathord{\ddagger}}}}
\newcommand*{\PSDoplus}{{\PSD^{\mathord{\oplus}}}}

\newcommand*{\SHDefplus}{{\SH^+_{\scriptscriptstyle \Def}}}
\newcommand*{\SHPSDplus}{{\SH^+_{\scriptscriptstyle \PSD}}}
\newcommand*{\Defoplus}{{\Def^{\mathord{\oplus}}}}

\newcommand*{\Def}{\mathit{Def}}
\newcommand*{\Pos}{\mathit{Pos}}
\newcommand*{\ScozzariShPSh}{\mathsf{Sh}^\mathsf{PSh}}
\newcommand*{\JonesSondG}{\mathsf{G}}

\newcommand*{\Con}{\mathit{Con}}

\newcommand*{\SHjplus}{{\SH^+_j}}

\newcommand*{\TSDone}{{\TSD_1}}

\newcommand*{\TSDtwo}{{\TSD_2}}

\newcommand*{\TSDj}{{\TSD_j}}
\newcommand*{\TSDk}{{\TSD_k}}
\newcommand*{\TSDn}{{\TSD_n}}

\newcommand*{\TSone}{{\TS_1}}
\newcommand*{\TStwo}{{\TS_2}}
\newcommand*{\TSi}{{\TS_{\!i}}}
\newcommand*{\TSj}{{\TS_{\!j}}}
\newcommand*{\TSh}{{\TS_h}}
\newcommand*{\TSk}{{\TS_k}}
\newcommand*{\TSn}{{\TS_n}}

\newcommand*{\rhoDef}{\mathop{\rho_{\scriptscriptstyle \Def}}}
\newcommand*{\rhoCon}{\mathop{\rho_{\scriptscriptstyle \Con}}}
\newcommand*{\rhoPS}{\mathop{\rho_{\scriptscriptstyle \PS}}}
\newcommand*{\rhoPSprime}{\mathop{\rho_{\scriptscriptstyle \PS'}}}
\newcommand*{\rhoPSD}{\mathop{\rho_{\scriptscriptstyle \PSD}}}

\newcommand*{\rhoTSDone}{\mathop{\rho_{\scriptscriptstyle \TSDone}}}
\newcommand*{\rhoTSDtwo}{\mathop{\rho_{\scriptscriptstyle \TSDtwo}}}
\newcommand*{\rhoTSDj}{\mathop{\rho_{\scriptscriptstyle \TSDj}}}
\newcommand*{\rhoTSDk}{\mathop{\rho_{\scriptscriptstyle \TSDk}}}

\newcommand*{\rhoTSone}{\mathop{\rho_{\scriptscriptstyle \TSone}}}
\newcommand*{\rhoTStwo}{\mathop{\rho_{\scriptscriptstyle \TStwo}}}
\newcommand*{\rhoTSj}{\mathop{\rho_{\scriptscriptstyle \TSj}}}
\newcommand*{\rhoTSh}{\mathop{\rho_{\scriptscriptstyle \TSh}}}
\newcommand*{\rhoTSk}{\mathop{\rho_{\scriptscriptstyle \TSk}}}

\newcommand*{\meet}{\wedge}

\newcommand*{\join}{\vee}
\newcommand*{\bigjoin}{\bigvee}

\newcommand*{\sqsslt}{\sqsubset}

\newtheorem{thm}{Theorem}
\newtheorem{cor}{Corollary}
\newtheorem{prop}{Proposition}
\newtheorem{lem}{Lemma}
\newtheorem{exmp}{Example}
\newtheorem{defn}{Definition}

\begin{document}
\title[Decomposing Non-Redundant Sharing]
      {Decomposing Non-Redundant Sharing \\ by Complementation}
\author[E. Zaffanella, P. M. Hill and R. Bagnara]
       {ENEA ZAFFANELLA \\
       Department of Mathematics,
       University of Parma,
       Italy
       \email{zaffanella@cs.unipr.it}
       \and PATRICIA M. HILL\thanks{This work was partly supported
                                 by EPSRC under grant GR/M05645.} \\
       School of Computing,
       University of Leeds,
       Leeds, U.K.
       \email{hill@comp.leeds.ac.uk}
       \and ROBERTO BAGNARA\thanks{The work of the first and third authors
                          has been partly supported by MURST project
                          ``Certificazione automatica di programmi
                            mediante interpretazione astratta.''} \\
       Department of Mathematics,
       University of Parma,
       Italy
       \email{bagnara@cs.unipr.it}
}
\maketitle

\begin{abstract}
Complementation,
the inverse of the reduced product operation,
is a technique for systematically finding
minimal decompositions of abstract domains.
Fil\'e and Ranzato advanced the state of the art by introducing
a simple method for computing a complement.
As an application, they considered the  extraction by complementation
of the pair-sharing
domain $\PS$ from the Jacobs and Langen's set-sharing domain $\SH$.
However, since the result of this operation was still $\SH$, 
they concluded that $\PS$ was too abstract for this.
Here, we show that the source of this result lies not with $\PS$
but with $\SH$ and, more precisely, with the redundant
information contained in $\SH$
with respect to ground-dependencies and pair-sharing.
In fact, a proper decomposition is obtained
if the non-redundant version of $\SH$, $\PSD$,
is substituted for $\SH$.
To establish the results for $\PSD$, we define a general schema
for subdomains of $\SH$ that includes $\PSD$ and $\Def$ as special cases.
This sheds new light on the structure of $\PSD$
and exposes a natural though unexpected connection between $\Def$ and $\PSD$.
Moreover, we substantiate the claim
that complementation \emph{alone} is not sufficient
to obtain \emph{truly minimal} decompositions of domains.
The right solution to this problem is to \emph{first} remove redundancies
by computing the quotient of the domain with respect to
the observable behavior, and only \emph{then} decompose it by complementation.
\end{abstract}


\section{Introduction}

Complementation \cite{CortesiFGPR97}, which is
the inverse of the well-known reduced product operation \cite{CousotC79},
can systematically obtain minimal decompositions of complex abstract domains.
It has been argued that these decompositions would be useful
in finding space saving representations for
domains and to simplify domain verification problems.

In \cite{FileR96}, Fil\'e and Ranzato presented a new method for computing
the complement, which is simpler than the original proposal
by Cortesi \etal{} \cite{CortesiFGPR95,CortesiFGPR97}
because it has the advantage that,
in order to compute the complement, only a relatively small number of elements
(namely the \emph{meet-irreducible} elements of the
reference domain) need be considered.
As an application of this method, the authors considered the
Jacobs and Langen's sharing domain \cite{JacobsL92}, $\SH$,
for representing properties of variables such as
groundness and sharing.
This domain captures the property of set-sharing.
Fil\'e and Ranzato illustrated their method by minimally decomposing
$\SH$ into three components;
using the words of the authors
\cite[Section~1]{FileR96}:
\begin{quotation}
``[$\ldots$] each representing one of the elementary properties
that coexist in the elements of \textit{Sharing}, and that are as follows:
(i) the ground-dependency information;
(ii) the pair-sharing information, or equivalently variable independence;
(iii) the set-sharing information, without variable independence and
ground-dependency.''
\end{quotation}
However, this decomposition did not use the usual domain $\PS$
for pair-sharing.
Fil\'e and Ranzato observed that the complement
of the pair-sharing domain $\PS$
with respect to $\SH$ is again $\SH$
and concluded that $\PS$ was too abstract to be extracted from $\SH$
by means of complementation.
Thus, in order to obtain their non-trivial decomposition of $\SH$,
they used a different (and somewhat unnatural) definition
for an alternative pair-sharing domain, called $\PS'$.
The nature of $\PS'$ and its connection with $\PS$ is examined more
carefully in Section~\ref{sec:discussion}.

We noticed that the reason why Fil\'e and Ranzato obtained this result
was not to be found in the definition of $\PS$,
which accurately represents the property of pair-sharing,
but in the use of the domain $\SH$
to capture the property of pair-sharing.
In~\cite{BagnaraHZ97b,BagnaraHZ01TCS}, it was observed that, for most
(if not all) applications, the property of interest is not set-sharing
but pair-sharing.
Moreover, it was shown that, for groundness
and pair-sharing, $\SH$ includes redundant elements.
By defining an upper closure operator $\rho$ that removed this redundancy,
a much smaller domain $\PSD$,
which was denoted $\SHm$ in \cite{BagnaraHZ97b},
was found that captured pair-sharing and groundness
with the same precision as $\SH$.
We show here that using the method given in \cite{FileR96},
but with this domain instead of $\SH$ as the reference domain,
a proper decomposition can be obtained even when considering
the natural definition of the pair-sharing domain $\PS$.
Moreover, we show that $\PS$ is \emph{exactly} one of the components
obtained by complementation of $\PSD$.
Thus the problem exposed by Fil\'e and Ranzato was, in fact, due to the
``information preserving'' property of complementation,
as any factorization obtained in this way is such that the reduced product
of the factors gives back the original domain.
In particular, any factorization of $\SH$ has to encode the
redundant information identified
in~\cite{BagnaraHZ97b,BagnaraHZ01TCS}.
We will show that such a problem disappears
when $\PSD$ is used as the reference domain.

Although the primary purpose of this work is to clarify the
decomposition of the domain $\PSD$, the formulation is sufficiently general
to apply to other properties that are captured by $\SH$.
The domain $\Pos$ of positive Boolean functions and
its subdomain $\Def$, the domain of \emph{definite} Boolean functions,
are normally used for capturing
groundness~\cite{ArmstrongMSS98}.
Each Boolean variable has the value \emph{true}
if the program variable it corresponds to is definitely bound to a ground term.
However, the domain $\Pos$
is isomorphic to $\SH$ via the mapping from
formulas in $\Pos$ to the set of complements of their models~\cite{CodishS98}.
This means that any general result regarding the structure of $\SH$
is equally applicable to $\Pos$ and its subdomains.

To establish the results for $\PSD$, we define a general schema
for subdomains of
$\SH$ that includes $\PSD$ and $\Def$ as special cases.
This sheds new light on the structure of the domain $\PSD$,
which is smaller but significantly more
involved than $\SH$.\footnote{For the well acquainted with the matter:
$\SH$ is a powerset and hence it is dual-atomistic;
this is not the case for $\PSD$.}
Of course, as we have used the more general schematic approach, we can
immediately derive (where applicable) corresponding results
for $\Def$ and $\Pos$.
Moreover, an interesting consequence of this work is the discovery
of a natural connection between the abstract domains
$\Def$ and $\PSD$.
The results confirm that $\PSD$ is, in fact, the ``appropriate''
abstraction of the set-sharing domain $\SH$ that has to be considered
when groundness and pair-sharing are the properties of interest.

The paper, which is an extended version of \cite{ZaffanellaHB99},
is structured as follows:
In Section~\ref{sec:preliminaries} we briefly recall the required notions
and notations, even though we assume general acquaintance with the topics of
lattice theory, abstract interpretation,
sharing analysis and groundness analysis.
Section~\ref{sec:sharing-domains} introduces the $\SH$ domain
and several abstractions of it.
The meet-irreducible elements of an important family of abstractions
of $\SH$ are identified in Section~\ref{sec:meet-irreducible}.
This is required in order to apply, in Section~\ref{sec:decomposition},
the method of Fil\'e and Ranzato to this family.
In Section~\ref{sec:discussion} we present some final remarks and
we explain what is, in our opinion, the lesson to be learned from this and
other related work. Section~\ref{sec:conclusion} concludes.

\section{Preliminaries}
\label{sec:preliminaries}

For any set $S$, $\wp(S)$ denotes the power set of $S$
and $\card S$ is the cardinality of $S$.

A \emph{preorder} `$\preceq$' over a set $P$ is a binary relation that is
reflexive and transitive.
If `$\preceq$' is also antisymmetric, then it is called \emph{partial order}.
A set $P$ equipped with a partial order `$\preceq$' is said
to be \emph{partially ordered}
and sometimes written $\langle P, \preceq\rangle$.
Partially ordered sets are also called \emph{posets}.

A poset $\langle P, \preceq\rangle$ is \emph{totally ordered}
with respect to `$\preceq$' if, for each $x, y \in P$,
either $x \preceq y$ or $y \preceq x$.
A subset $S$ of a poset $\langle P, \preceq\rangle$
is a \emph{chain} if it is totally ordered
with respect to `$\preceq$'.

Given a poset $\langle P, \preceq\rangle$ and $S \sseq P$, $y \in P$ is
an \emph{upper bound} for $S$ if and only if $x \preceq y$ for each $x \in S$.
An upper bound $y$ for $S$ is a \emph{least upper bound} (or $\lub$)
of $S$ if and only if, for every upper bound $y'$ for $S$,
$y \preceq y'$.
The $\lub$, when it exists, is unique. In this case we write $y = \lub S$.
\emph{Lower bounds} and \emph{greatest lower bounds} (or $\glb$)
are defined dually.

A poset $\langle L, \preceq\rangle$ such that, for each $x, y \in L$,
both $\lub \{ x, y \}$ and $\glb \{ x, y \}$ exist, is called a
\emph{lattice}. In this case, $\lub$ and $\glb$ are also called, respectively,
the \emph{join} and the \emph{meet} operations of the lattice.
A \emph{complete lattice} is a lattice $\langle L, \preceq\rangle$
such that every subset of $L$ has both a least upper bound
and a greatest lower bound. The \emph{top} element of
a complete lattice $L$, denoted by $\top$, is such that
$\top \in L$ and $\forall x \in L \itc x \preceq \top$.
The \emph{bottom} element of $L$, denoted by $\bot$,
is defined dually.

As an alternative definition, a lattice is
an algebra $\langle L, \meet, \join \rangle$ such that
$\meet$ and $\join$ are two binary operations
over $L$ that are commutative, associative, idempotent, and satisfy the
following \emph{absorption laws}, for each $x, y \in L$:
$x \meet (x \join y) = x$ and
$x \join (x \meet y) = x$.

The two definitions of lattice are equivalent.
This can be seen by defining:
\[
  x \preceq y \quad\defiff\quad x \meet y = x \quad\defiff\quad x \join y = y
\]
and
\begin{align*}
  \glb \{ x, y \} &\defeq x \meet y, \\
  \lub \{ x, y \} &\defeq x \join y.
\end{align*}
The existence of an isomorphism between the two lattices
$L_1$ and $L_2$ is denoted by $L_1 \equiv L_2$.

A monotone and idempotent self-map $\fund{\rho}{P}{P}$
over a poset $\langle P, \preceq\rangle$ is
called a \emph{closure operator} (or \emph{upper closure operator})
if it is also \emph{extensive}, namely
\[
  \forall x \in P \itc x \preceq \rho(x).
\]
Each upper closure operator $\rho$ over a complete lattice $C$
is uniquely determined by the set of its fixpoints, that is, by
its image
\begin{align*}
  \rho(C)
    &\defeq
      \bigl\{\, \rho(x) \bigm| x \in C \,\bigr\}. \\
\intertext{%
We will often denote upper closure operators by their images.
The set of all upper closure operators over a complete lattice $C$,
denoted by $\uco(C)$, forms a complete lattice ordered as follows:
if $\rho_1, \rho_2 \in \uco(P)$,
$\rho_1 \sqsseq \rho_2$ if and only if $\rho_2(C) \sseq \rho_1(C)$.
The \emph{reduced product} of two elements $\rho_1$ and $\rho_2$
of $\uco(C)$ is denoted by $\rho_1 \ucomeet \rho_2$ and defined as
}
  \rho_1 \ucomeet \rho_2
    &\defeq
      \glb\{\rho_1,\rho_2\}.
\end{align*}
For a more detailed introduction to closure operators,
the reader is referred to \cite{GHK+80}.

A complete lattice $C$ is \emph{meet-continuous} if for any chain $Y \sseq C$
and each $x \in C$,
\[
  x \meet \bigl(\bigjoin Y\bigr) = \bigjoin_{y \in Y} (x \meet y).
\]
Most domains for abstract interpretation~\cite{CortesiFGPR97}
and, in particular, all the domains considered in this paper
are meet-continuous.

Assume that $C$ is a meet-continuous lattice.
Then the inverse of the reduced product operation, called
\emph{weak relative pseudo-complement},
is well defined and given as follows.
Let $\rho, \rho_1 \in \uco(C)$ be such that $\rho \sqsseq \rho_1$.
Then
\[
  \rho \sim \rho_1
    \defeq
      \lub\{\, \rho_2 \in \uco(C) \mid \rho_1 \ucomeet \rho_2 = \rho \,\}.
\]
Given $\rho \in \uco(C)$,
the \emph{weak pseudo-complement}
(or, by an abuse of terminology now customary in the field
of Abstract Interpretation, simply \emph{complement}) of $\rho$
is denoted by $\idc \sim \rho$, where $\idc$ is the identity over $C$.
Let $D_i \defeq \rho_{D_i}(C)$
with $\rho_{D_i} \in \uco(C)$ for $i = 1$, \dots,~$n$.
Then $\{\, D_i \mid 1\leq i\leq n \,\}$ is a \emph{decomposition for $C$}
if $C = D_1 \ucomeet \cdots \ucomeet D_n$.
The decomposition is also called \emph{minimal} if,
for each $k \in \Nset$ with $1 \leq k \leq n$
and each $E_k \in \uco(C)$,
$D_k \sqsslt E_k$ implies
\[
  C
    \sqsslt
      D_1 \ucomeet \cdots \ucomeet D_{k-1}
        \ucomeet E_k \ucomeet D_{k+1}
          \ucomeet \cdots \ucomeet D_n.
\]

Assume now that $C$ is a complete lattice.
If $X \sseq C$,
then $\Moore(X)$ denotes the \emph{Moore completion of $X$},
namely,
\[
  \Moore(X) \defeq \bigl\{\, \bigwedge Y \bigm| Y \sseq X \,\bigm\}.
\]
We say that \emph{$C$ is meet-generated by $X$} if $C = \Moore(X)$.
An element $x \in C$ is \emph{meet-irreducible} if
\[
  \forall y, z \in C
    \itc \bigl((x = y \meet z) \implies (x = y \text{ or } x = z)\bigr).
\]
The set of meet-irreducible elements of a
complete lattice $C$ is denoted by $\MI(C)$.
Note that $\top \in \MI(C)$.
An element $x \in C$ is a \emph{dual-atom} if $x \neq \top$ and,
for each $y \in C$, $x \leq y < \top$ implies $x = y$.
The set of dual-atoms is denoted by $\dAtoms(C)$.
Note that $\dAtoms(C) \sslt \MI(C)$.
The domain $C$ is \emph{dual-atomistic} if $C = \Moore\bigl(\dAtoms(C)\bigr)$.
Thus, if $C$ is dual-atomistic,
$\MI(C) = \{\top\} \union \dAtoms(C)$.
The following result
holds~\cite[Theorem~4.1]{FileR96}.
\begin{thm}
\label{thm:FileR96}
If $C$ is meet-generated by $\MI(C)$ then $\uco(C)$ is
pseudo-com\-ple\-mented and for any $\rho \in  \uco(C)$
\[
  \idc \sim \rho = \Moore\bigl(\MI(C) \setdiff \rho(C)\bigr).
\]
\end{thm}
Another interesting result is the following
\cite[Corollary~4.5]{FileR96}.
\begin{thm}
\label{thm: dual-atoms}
If $C$ is dual-atomistic then $\uco(C)$ is pseudo-complemented
and for any $\rho \in \uco(C)$
\[
  \idc \sim \rho = \Moore\bigl(\dAtoms(C) \setdiff \rho(C)\bigr).
\]
\end{thm}

Let $\Vars$ be a denumerable set of variables.
For any syntactic object $o$, $\vars(o)$ denotes
the set of variables occurring in $o$.
Let $\Terms$ be the set of first-order terms over $\Vars$.
If $x \in \Vars$ and $t \in \Terms \setdiff \{ x \}$,
then $x \mapsto t$ is called a \emph{binding}.
A \emph{substitution} is a total function $\fund{\sigma}{\Vars}{\Terms}$
that is the identity almost everywhere.
Substitutions are denoted by the set of their bindings,
thus a substitution $\sigma$ is identified with the (finite) set
\[
  \bigl\{\, x \mapsto \sigma(x) \bigm| x \neq \sigma(x) \,\bigr\}.
\]
If $t \in \Terms$, we write $t \sigma$ to denote $\sigma(t)$.
A substitution $\sigma$ is \emph{idempotent} if, for all $t\in\Terms$,
we have $t\sigma\sigma=t\sigma$. The set of all idempotent substitutions
is denoted by $\Subst$.

It should be stressed that this restriction to idempotent substitutions
is provided for presentation purposes only.
In particular, it allows for a straight comparison of our work
with respect to other works appeared in the literature.
However, the results proved in this paper do not rely on the idempotency
of substitutions and are therefore applicable also when considering
substitutions in \emph{rational solved form} \cite{Colmerauer82,Colmerauer84}.
Indeed, we have proved in \cite{HillBZ98b} that
the usual abstract operations defined on the domain $\SH$,
approximating concrete unification over finite trees,
also provide a correct approximation of concrete unification
over a domain of rational trees.

\section{The \textsf{Sharing} Domains}
\label{sec:sharing-domains}

In order to provide a concrete meaning to the elements of the set-sharing
domain of D.~Jacobs and A.~Langen~\cite{JacobsL89,Langen90th,JacobsL92},
a knowledge of the finite set $\VI \sslt \Vars$ of variables of interest
is required.
For example, in the Ph.D.~thesis of Langen~\cite{Langen90th} this set
is implicitly defined, for each clause being analyzed, as the finite set
of variables occurring in that clause.
A clearer approach has been introduced in~\cite{CortesiFW94,CortesiFW98}
and also adopted in~\cite{BagnaraHZ97b,BagnaraHZ01TCS,CortesiF99},
where the set of variables of interest is given explicitly
as a component of the abstract domain.
During the analysis process, this set is \emph{elastic}.
That is, it expands (e.g., when solving clause's bodies)
and contracts (e.g., when abstract descriptions are projected
onto the variables occurring in clause's heads).
This technique has two advantages: 
first, a clear and unambiguous description of those semantic operators
that modify the set of variables of interest is provided;
second, the definition of the abstract domain is completely independent
from the particular program being analyzed.
However, since at any given time the set of variables of interest is fixed,
we can simplify the presentation by consistently denoting this set by $\VI$.
Therefore, in this paper all the abstract domains defined are restricted
to a fixed set of variables of interest $\VI$ of finite cardinality $n$;
this set is not included explicitly in the representation
of the domain elements;
also, when considering abstract semantic operators having some arguments
in $\Subst$, such as the abstract $\mgu$, the considered substitutions
are always taken to have variables in $\VI$.
We would like to emphasize that
this is done for ease of presentation only:
the complete definition of both the domains and the semantic operators
can be immediately derived from those given, e.g.,
in~\cite{BagnaraHZ97b,BagnaraHZ01TCS}.
Note that other solutions are possible; we refer the interested reader
to~\cite[Section~7]{CortesiFW96}
and~\cite[Section~10]{Scozzari01TCS},
where this problem is discussed in the context of groundness analysis.

\subsection{The Set-sharing Domain $\SH$}

\begin{defn} \summary{(The \emph{set-sharing} domain $\SH$.)}
The domain $\SH$ is given by
\begin{align*}
 \SH &\defeq \wp(\SG), \\
\intertext{%
where the set of \emph{sharing-groups} $\SG$ is given by
}
  \SG
    &\defeq
      \wp(\VI) \setdiff \{ \emptyset \}.
\end{align*}
$\SH$ is partially ordered by set inclusion so that
the $\lub$ is given by set union and the $\glb$ by set intersection.
\end{defn}
Note that, as we are adopting the upper closure operator approach to
abstract interpretation, all the domains we define here
are ordered by subset inclusion.
As usual in the field of abstract interpretation,
this ordering provides a formalization of precision
where the less precise domain elements are those occurring higher
in the partial order.
Thus, more precise elements contain less sharing groups.

Since $\SH$ is a power set, $\SH$ is dual-atomistic and
\[
  \dAtoms(\SH)
    =
      \bigl\{\,
        \SG \setdiff \{S\}
      \bigm|
         S \in \SG
      \,\bigr\}.
\]

In all the examples in this paper, the elements of $\SH$
are written in a simplified notation, omitting the inner braces.
For instance, the set
\[
  \bigl\{ \{x\}, \{x,y\}, \{x,z\}, \{x,y,z\}\bigr\}
\]
would be written simply as
\[
   \{ x, xy, xz, xyz \}.
\]

\begin{exmp}
\label{ex: three-vars-datoms}
Suppose $\VI = \{ x, y, z \}$.
Then the seven dual-atoms of $\SH$ are:
\begin{align*}
  &\left.
    \begin{aligned}
      s_1 &= \{\phantom{x,{}} y,z,xy,xz,yz,xyz\},\\
      s_2 &= \{x,\phantom{y,{}} z,xy,xz,yz,xyz\},\\
      s_3 &= \{x,y,\phantom{z,{}} xy,xz,yz,xyz\},
    \end{aligned}
  \right\} &&\text{ these lack a singleton}; \\
  &\left.
    \begin{aligned}
      s_4 &= \{x,y,z,\phantom{xy,{}} xz,yz,xyz\},\\
      s_5 &= \{x,y,z,xy,\phantom{xz,{}} yz,xyz\},\\
      s_6 &= \{x,y,z,xy,xz,\phantom{yz,{}} xyz\},\\
    \end{aligned}
  \right\} &&\text{ these lack a pair}; \\
  &\left.
    \begin{aligned}
      s_7 &= \{x,y,z,xy,xz,yz\phantom{{},xyz}\},
    \end{aligned}
  \right.  &&\text{ this lacks $\VI$.}
\end{align*}
\end{exmp}
The meet-irreducible elements of $\SH$ are
$s_1$,\dots,~$s_7$, and the top element $\SG$.

\begin{defn} \summary{(Operations over $\SH$.)}
\label{def:abs-funcs}
The function
$\fund{\bin}{\SH\times\SH}{\SH}$,
called \emph{binary union},
is given, for each $\sh_1, \sh_2 \in \SH$, by
\begin{align*}
  \bin(\sh_1, \sh_2)
    &\defeq
      \{\,
        S_1 \union S_2
      \mid
        S_1 \in \sh_1,
        S_2 \in \sh_2
      \,\}. \\
\intertext{%
The \emph{star-union}  function
$\fund{(\cdot)^\star}{\SH}{\SH}$
is given,
for each $\sh \in \SH$,
by
}
  \sh^\star
    &\defeq
      \Bigl\{\,
        S \in \SG
      \Bigm|
            \exists \sh' \sseq \sh
              \st S = \bigunion \sh'
      \,\Bigr\}.\\
\intertext{%
The \emph{$j$-self-union} function $\fund{(\cdot)^j}{\SH}{\SH}$ is given,
for each $j \geq 1$ and $\sh \in \SH$,
by
}
  \sh^j
    &\defeq
      \Bigl\{\,
        S \in \SG
      \Bigm|
        \exists \sh' \sseq \sh
          \st \Bigl(\card \sh' \leq j, S = \bigunion \sh'\Bigr)
      \,\Bigr\}. \\
\intertext{%
The extraction of the
\emph{relevant component} of an element of $\SH$
with respect to a subset of $\VI$
is encoded by the function
$\fund{\rel}{\wp(\VI)\times\SH}{\SH}$
given, for each $V \sseq \VI$ and each $\sh \in \SH$, by
}
  \rel(V, \sh)
    &\defeq
      \{\, S \in \sh \mid S \inters V \neq \emptyset \,\}.\\
\intertext{%
The function $\amgu$ captures the effects of a binding $x \mapsto t$
on an element of $\SH$.
Let $\sh \in \SH$,
$v_x = \{x\}$, $v_t = \vars(t)$, and $v_{xt} = v_x \union v_t$.
Then
}
  \amgu(\sh, x \mapsto t)
    &\defeq
      \bigl(\sh \setdiff (\rel(v_{xt},\sh)\bigr)
        \union \bin\bigl(\rel(v_x,\sh)^\star, \rel(v_t,\sh)^\star\bigr). \\
\intertext{%
  We also define the extension
  \(
    \fund{\amgu}{\SH\times\Subst}{\SH}
  \)
  by
}
  \amgu(\sh, \emptyset)
    &\defeq
      \sh, \\
  \amgu\bigl(\sh, \{x \mapsto t\} \union \sigma\bigr)
    &\defeq
      \amgu\bigl(\amgu(\sh, x \mapsto t),
                 \sigma \setdiff \{x \mapsto t\}\bigr).\\
\intertext{%
The function $\fund{\proj}{\SH\times\wp(\VI)}{\SH}$
that \emph{projects} an element of $\SH$ onto a subset $V \sseq \VI$
of the variables of interest is given, for each $\sh \in \SH$, by
}
  \proj(\sh, V)
    &\defeq
      \{\, S \inters V \mid S \in \sh, S \inters V \neq \emptyset \,\}
      \union
      \bigl\{\, \{ x \} \bigm| x \in \VI \setdiff V \,\bigr\}.
\end{align*}
\end{defn}

Together with $\lub$, the functions $\proj$ and $\amgu$ 
are the key operations that make the abstract domain $\SH$
suitable for computing static approximations of
the substitutions generated by the execution of logic programs.
These operators can be combined with simpler ones
(e.g., consistent renaming of variables)
so as to provide a complete definition of the abstract semantics.
Also note that these three operators have been proved to be
the \emph{optimal approximations}
of the corresponding concrete operators~\cite{CortesiF99}.
The $j$-self-union operator defined above is new. 
We show later when it 
may safely replace the star-union operator.
Note that, letting $j = 1$, $2$, and~$n$, we have
$\sh^1 = \sh$, $\sh^2 = \bin(\sh,\sh)$,
and, as $\card \VI = n$, $\sh^n = \sh^\star$.

\subsection{The Tuple-Sharing Domains}

To provide a general characterization of domains such as the groundness and
pair-sharing domains contained in $\SH$,
we first identify the sets of elements that have the same cardinality.

\pagebreak
\begin{defn} \summary{(Tuples of cardinality $k$.)}
For each $k \in \Nset$ with $1\leq k \leq n$, the overloaded functions
$\fund{\tuples_k}{\SG}{\SH}$ and
$\fund{\tuples_k}{\SH}{\SH}$
are defined as
\begin{align*}
  \tuples_k(S)
    &\defeq
      \bigl\{\,
        T \in \wp(S)
      \bigm|
        \card T =k
      \,\bigr\}, \\
  \tuples_k(\sh)
    &\defeq
      \bigunion
        \bigl\{\,
          \tuples_k(S')
        \bigm|
           S' \in \sh
        \,\bigr\}. \\
\intertext{%
In particular, if $S \in \SG$ and $\sh \in\SH$, let
}
  \pairs(S) &\defeq \tuples_2(S),\\
  \pairs(\sh) &\defeq \tuples_2(\sh).
\end{align*}
\end{defn}

The usual domains that represent groundness and pair-sharing information
will be shown to be special cases of the following more general domain.

\begin{defn} \summary{(The \emph{tuple-sharing} domains $\TSk$.)}
\label{def: tuple-domain}
For each $k \in \Nset$ such that $1\leq k \leq n$,
the function $\fund{\rhoTSk}{\SH}{\SH}$ is
defined as
\begin{align*}
  \rhoTSk(\sh)
    &\defeq
      \bigl\{\,
        S \in \SG
      \bigm|
        \tuples_k(S) \sseq \tuples_k(\sh)
      \,\bigr\} \\
\intertext{%
and, as $\rhoTSk \in\uco(\SH)$, it induces the lattice
}
  \TSk &\defeq  \rhoTSk(\SH).
\end{align*}
\end{defn}
Note that $\rhoTSk\bigl(\tuples_k(\sh)\bigr) = \rhoTSk(\sh)$ and that there
is a one to one correspondence between $\TSk$ and
$\wp\bigl(\tuples_k(\VI)\bigr)$.
The isomorphism is given by the functions
$\fund{\tuples_k}{\TSk}{\wp\bigl(\tuples_k(\VI)\bigr)}$ and
$\fund{\rhoTSk}{\wp\bigl(\tuples_k(\VI)\bigr)}{\TSk}$.
Thus the domain $\TSk$ is the smallest domain
that can represent properties characterized by sets of variables of
cardinality $k$.
We now consider the tuple-sharing domains
for the cases when $k=1$, $2$, and~$n$.
\begin{defn} \summary{(The \emph{groundness} domain $\Con$.)}
The upper closure operator $\fund{\rhoCon}{\SH}{\SH}$
and the corresponding domain $\Con$ are defined as
\begin{align*}
  \rhoCon
    &\defeq \rhoTSone, \\
  \Con
    &\defeq
      \TSone(\SH) = \rhoCon(\SH).
\end{align*}
\end{defn}
This domain, which represents groundness information,
is isomorphic to a domain of conjunctions of Boolean variables.
The isomorphism $\tuples_1$ maps each element of
$\Con$ to the set of variables that are possibly non-ground.
From the domain $\tuples_1(\Con)$, by set complementation,
we obtain the classical domain $\JonesSondG$~\cite{JonesS87}
for representing the set of variables that are definitely ground
(so that we have $\TSone \defeq \Con \equiv \JonesSondG$).

\pagebreak
\begin{defn} \summary{(The \emph{pair-sharing} domain $\PS$.)}
The upper closure operator $\fund{\rhoPS}{\SH}{\SH}$
and the corresponding domain $\PS$ are defined as
\begin{align*}
  \rhoPS
    &\defeq \rhoTStwo, \\
  \PS
    &\defeq
      \TStwo(\SH) = \rhoPS(\SH).
\end{align*}
\end{defn}
This domain represents
pair-sharing information and the isomorphism $\tuples_2$ maps each element of
$\PS$ to the set of pairs of variables that may be
bound to terms that share a common variable.
The domain for representing variable independence
can be obtained by set complementation.

Finally, in the case when $k=n$ we have a domain consisting of just
two elements:
\[
  \TSn = \bigl\{ \SG, \SG \setdiff \{\VI\} \bigr\}.
\]
Note that the bottom of $\TSn$ differs from the top element $\SG$
only in that it lacks the sharing group $\VI$.
There is no intuitive reading for the information encoded by this
element: it describes all but those substitutions $\sigma \in \Subst$
such that
\(
  \biginters
    \bigl\{\,
      \vars(x\sigma)
    \bigm|
      x \in \VI
    \,\bigr\}
      \neq \emptyset
\).

Just as for $\SH$, the domain $\TSk$ (where $1 \leq k \leq n$)
is dual-atomistic and:
\begin{align*}
 \dAtoms(\TSk) &=
  \Bigl\{\,
    \bigl( \SG \setdiff \{\, U \in \SG \mid T \sseq U\,\} \bigr)
  \Bigm|
    T \in \tuples_k(\VI)
  \,\Bigr\}.\\
\intertext{%
Thus we have
}
\dAtoms(\Con) &=
  \Bigl\{\,
    \bigl( \SG \setdiff \{\, U \in \SG \mid x \in U\,\} \bigr)
  \Bigm|
    x\in \VI
  \,\Bigr\},\\
\dAtoms(\PS) &=
  \Bigl\{\,
    \bigl( \SG \setdiff \{\, U \in \SG \mid x,y \in U\,\} \bigr)
  \Bigm|
    x, y \in \VI, x \neq y
  \,\Bigr\}.
\end{align*}

\begin{exmp}
\label{ex: Con-PS-datoms}
Consider Example~\ref{ex: three-vars-datoms}.
Then the dual-atoms of $\Con$ are
\begin{align*}
      r_1 =  s_1 \inters s_4 \inters s_5 \inters s_7
          &= \{\phantom{x,{}} y,z,\phantom{xy,{}} \phantom{xz,{}} yz\},\\
      r_2 =  s_2 \inters s_4 \inters s_6 \inters s_7
          &= \{x, \phantom{y,{}} z,\phantom{xy,{}} xz \phantom{,{}yz}\},\\
      r_3 =  s_3 \inters s_5 \inters s_6 \inters s_7
          &= \{x,y, \phantom{z,{}} xy \phantom{,{}xz} \phantom{,{}yz}\};\\
\intertext{%
the dual-atoms of $\PS$ are
}
      m_1 =  s_4 \inters s_7
          &= \{x,y,z,\phantom{xy,{}} xz, yz\},\\
      m_2 =  s_5 \inters s_7
          &= \{x,y,z,xy,  \phantom{xz,{}} yz\},\\
      m_3 =  s_6 \inters s_7
          &= \{x,y,z, xy, xz \phantom{,{}yz}\}.
\end{align*}
\end{exmp}

It can be seen from the dual-atoms that,
for each $j = 1$, \dots,~$n$, where $j \neq k$,
the precision of the information encoded by domains
$\TSj$ and $\TSk$ is not comparable.
Also, we note that, if $j<k$, then $\rhoTSj(\TSk) = \{\SG\}$
and  $\rhoTSk(\TSj) = \TSj$.

\subsection{The Tuple-Sharing Dependency Domains}

We now need to define domains that capture the propagation of
groundness and pair-sharing; in particular,
the dependency of these properties on the further
instantiation of the variables.
In the same way as with $\TSk$ for $\Con$
and $\PS$, we first define a general subdomain $\TSDk$ of $\SH$.
This must be safe with respect to the tuple-sharing property
represented by $\TSk$ when performing the usual abstract operations.
This was the motivation behind the introduction
in~\cite{BagnaraHZ97b,BagnaraHZ01TCS}
of the pair-sharing dependency domain $\PSD$.
We now generalize this for tuple-sharing.

\begin{defn} \summary{The \emph{tuple-sharing dependency} domain ($\TSDk$.)}
\label{def:rhoSTk}
For each $k$ where $1\leq k \leq n$,
the function $\fund{\rhoTSDk}{\SH}{\SH}$ is defined as
\begin{multline*}
  \rhoTSDk(\sh) \\
    \defeq
        \Bigl\{\,
          S \in \SG
        \Bigm|
          \forall T \sseq S
            \itc \card T <k
              \implies
                S = \bigunion \{\,
                                U \in \sh
                              \mid
                                T \sseq U \sseq S
                              \,\}
        \,\Bigr\},
\end{multline*}
and, as $\rhoTSDk \in\uco(\SH)$,
it induces the \emph{tuple-sharing dependency} lattice
\[
  \TSDk \defeq  \rhoTSDk(\SH).
\]
\end{defn}
It follows from the definitions that the domains $\TSDk$ form a strict chain.
\begin{prop}
\label{prop:TSDk-chain}
For $j, k \in \Nset$ with $1\leq j < k \leq n$, we have
\(
  \TSDj \sslt \TSDk.
\)
\end{prop}
Moreover, $\TSDk$ is not less precise than $\TSk$.
\begin{prop}
\label{prop:TSDk-vs-TSk}
For $k \in \Nset$ with $1\leq k \leq n$, we have
\(
  \TSk \sseq \TSDk.
\)
Furthermore, if $n > 1$ then
\(
  \TSk \sslt \TSDk.
\)
\end{prop}
As an immediate consequence
of Propositions~\ref{prop:TSDk-chain} and~\ref{prop:TSDk-vs-TSk}
we have that that $\TSDk$ is not less precise than
$\TSone \ucomeet \cdots \ucomeet \TSk$.
\begin{cor}
\label{cor:TSDk-vs-TSj}
For $j, k \in \Nset$ with $1 \leq j \leq k \leq n$, we have
\(
  \TSj \sseq \TSDk.
\)
\end{cor}
It also follows from the definitions that, for the $\TSDk$ domain,
the star-union operator can be replaced by the $k$-self-union operator.
\begin{prop}
\label{pro: k-self-union}
For $1\leq k \leq n$, we have
\(
  \rhoTSDk\bigl(\sh^k\bigr) = \sh^\star.
\)
\end{prop}

We now instantiate the tuple-sharing dependency domains
for the cases when $k=1$, $2$, and~$n$.

\pagebreak
\begin{defn} \summary{(The \emph{ground dependency} domain $\Def$.)}
The domain $\Def$ is induced by the upper closure operator
$\fund{\rhoDef}{\SH}{\SH}$.
They are defined as
\begin{align*}
  \rhoDef
    &\defeq \rhoTSDone, \\
  \Def
    &\defeq
        \TSDone =  \rhoDef(\SH).
\end{align*}
\end{defn}
By Proposition~\ref{pro: k-self-union}, we have, for all $\sh \in \SH$,
$\rhoTSDone(\sh) = \sh^\star$ so that
$\TSDone$ is a representation of the domain $\Def$
used for capturing groundness.
It also provides evidence for the fact that the computation of the star-union
is not needed for the elements in $\Def$.
\begin{defn} \summary{(The \emph{pair-sharing dependency} domain $\PSD$.)}
The upper closure operator $\fund{\rhoPSD}{\SH}{\SH}$
and the corresponding domain $\PSD$ are defined as
\begin{align*}
  \rhoPSD
    &\defeq \rhoTSDtwo, \\
  \PSD
    &\defeq
        \TSDtwo =  \rhoPSD(\SH).
\end{align*}
\end{defn}
Then, it follows
from~\cite[Theorem~7]{BagnaraHZ97b}
that $\PSD$ corresponds to the domain $\SHm$
defined for capturing pair-sharing.
By Proposition~\ref{pro: k-self-union} we have, for all $\sh \in \SH$,
that $\rhoPSD(\sh^2) = \sh^\star$,
so that, for elements in $\PSD$,
the star-union operator $\sh^\star$ can be replaced by the
$2$-self-union $\sh^2 = \bin(\sh, \sh)$ without any loss of precision.
This was also proved in~\cite[Theorem~11]{BagnaraHZ97b}.
Furthermore, Corollary~\ref{cor:TSDk-vs-TSj} confirms the observation
made in~\cite{BagnaraHZ97b}
that $\PSD$ also captures groundness.

\begin{figure}
\def\objectstyle{\displaystyle}
\xymatrix@!C=1cm@!R=1cm{
  \TSone = \Con & \TStwo = \PS & \TS_3 \ar@{.}[rr] & & \TS_{n-1} & \TSn \\
  & \bigucomeet_{i=1}^2 \TSi \ar@{-}[ul] \ar@{-}[u] \\
  \TSDone = \Def \ar@{-}[uu] & & \bigucomeet_{i=1}^3 \TSi \ar@{-}[ul] \ar@{-}[uu] \\
  & \TSDtwo = \PSD \ar@{-}[uu] \ar@{-}[ul] \\
  & & \TSD_3 \ar@{-}[uu] \ar@{-}[ul] & & \bigucomeet_{i=1}^{n-1} \TSi \ar@{-}[uuuu] \ar@{.}[ulul] \\
  & & & & & \\
  & & & & \TSD_{n-1} \ar@{-}[uu] \ar@{.}[ulul] \\
  & & & & & \TSDn = \bigucomeet_{i=1}^n \TSi = \SH \ar@{-}[uuuuuuu] \ar@{-}[ul] \ar@{-}[uuul] \\
}
\caption{The set-sharing domain $\SH$ and some of its abstractions.}
\label{fig:relations-between-sharing-domains}
\end{figure}
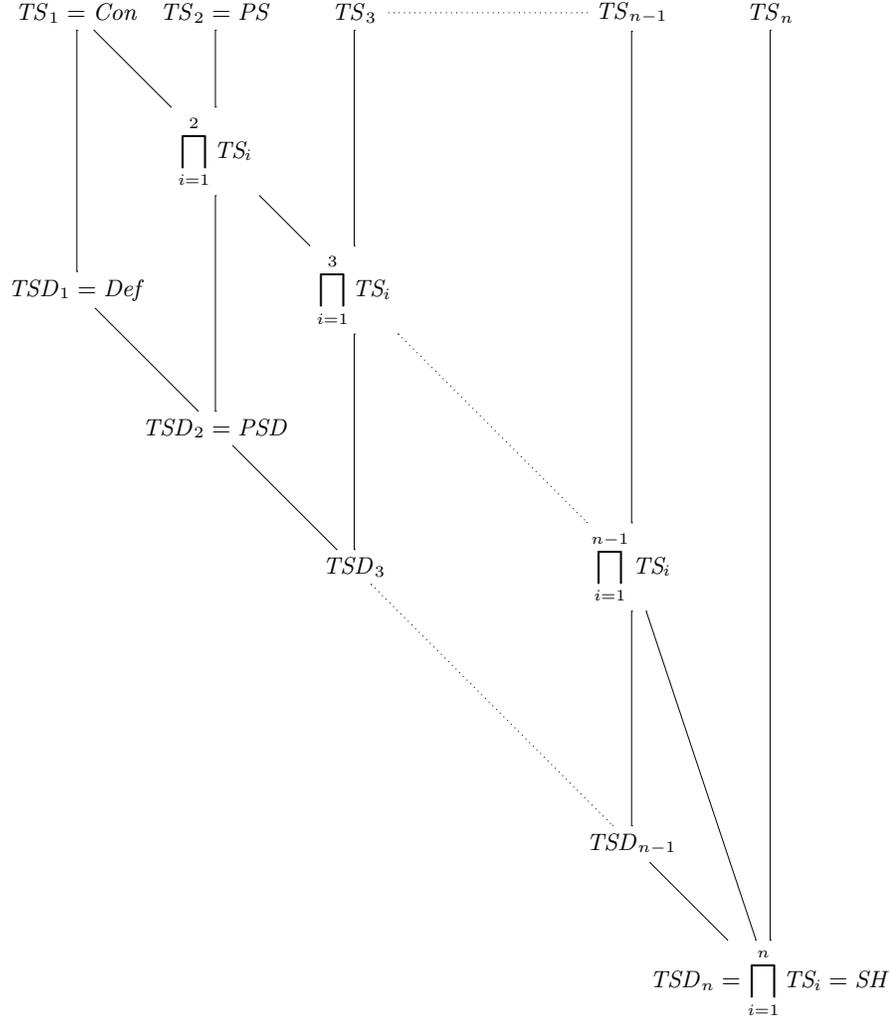

Finally, letting $k=n$, we observe that $\TSDn = \SH$.
Figure~\ref{fig:relations-between-sharing-domains} summarizes
the relations between the tuple-sharing and the tuple-sharing dependency
domains.

As already discussed at the start of this section,
the set of variables of interest $\VI$ is fixed and, to simplify
the notation, omitted.
In~\cite{BagnaraHZ97b,BagnaraHZ01TCS} the domains $\SSl$ and $\SSm$
(corresponding to $\SH$ and $\PSD$, respectively)
are instead obtained by explicitly adding to each domain element
a new component, representing the set of variables of interest.
It is shown that $\SSm$ is as good as $\SSl$ for both representing and
propagating pair-sharing and it is also proved that any weaker domain
does not satisfy these properties,
so that $\SSm$ is the quotient \cite{CortesiFW94,CortesiFW98} of $\SSl$
with respect to the pair-sharing property $\PS$.

We now generalize and strengthen the results
in~\cite{BagnaraHZ97b,BagnaraHZ01TCS}
and show that, for each $k \in \{1,\ldots,n\}$,
$\TSDk$ is the quotient of $\SH$
with respect to the reduced product $\TSone \ucomeet \cdots \ucomeet \TSk$.
These results are proved at the end of this section.
\begin{thm}
\label{thm: no-loss-ops}
Let $\sh_1, \sh_2 \in \SH$ and $1\leq k \leq n$.
If $\rhoTSDk(\sh_1) = \rhoTSDk(\sh_2)$ then,
for each $\sigma \in \Subst$, each $\sh' \in \SH$,
and each $V \in \wp(\VI)$,
\begin{align*}
     \rhoTSDk\bigl(\amgu(\sh_1, \sigma)\bigr)
  &= \rhoTSDk\bigl(\amgu(\sh_2, \sigma)\bigr), \\
     \rhoTSDk(\sh' \union \sh_1)
  &= \rhoTSDk(\sh' \union \sh_2), \\
     \rhoTSDk\bigl(\proj(\sh_1, V)\bigr)
  &= \rhoTSDk\bigl(\proj(\sh_2, V)\bigr).
\end{align*}
\end{thm}

\begin{thm}
\label{thm: smallest-domain-for-ops}
Let $1\leq k \leq n$
For each $\sh_1, \sh_2 \in \SH$,
$\rhoTSDk(\sh_1) \neq \rhoTSDk(\sh_2)$
implies
\[
    \exists \sigma \in \Subst, \exists j \in \{1,\ldots,k\} \st
      \rhoTSj\bigl(\amgu(\sh_1, \sigma)\bigr)
        \neq \rhoTSj\bigl(\amgu(\sh_2,\sigma)\bigr).
\]
\end{thm}

\subsection{Proofs of Theorems~\ref{thm: no-loss-ops}
and~\ref{thm: smallest-domain-for-ops}}

In what follows we use the fact that
$\rhoTSDk$ is an upper closure operator so that,
for each $\sh_1, \sh_2 \in \SH$,
\begin{align}
\label{eq-emi-rho}
  \sh_1 \sseq \rhoTSDk(\sh_2)
    &\iff
      \rhoTSDk(\sh_1) \sseq \rhoTSDk(\sh_2). \\
\intertext{%
In particular, since $ (\cdot)^\star = \rhoTSDone$, we have
}
\label{eq-emi-star}
  \sh_1 \sseq \sh_2^\star
    &\iff
      \sh_1^\star \sseq \sh_2^\star.
\end{align}

\begin{lem}
\label{lem:addit}
For each $\sh \in \SH$ and each $V \in \wp(\VI)$,
\[
  \rhoTSDk(\sh) \setdiff \rel\bigl(V, \rhoTSDk(\sh)\bigr)
    = \rhoTSDk\bigl(\sh \setdiff \rel(V,\sh)\bigr).
\]
\end{lem}
\begin{proof*}
By Definition~\ref{def:rhoSTk},
\begin{align*}
  S \in &\rhoTSDk\bigl(\sh \setdiff \rel(V, \sh)\bigr) \\
    &\iff
      \forall T \sseq S
        \itc
          \Bigl(
            \card T < k
              \implies
                S = \bigunion \bigl\{\,
                                U \in \sh \setdiff \rel(V, \sh)
                              \bigm|
                                T \sseq U \sseq S
                              \,\bigr\}
          \Bigr) \\
    &\iff
      \forall T \sseq S
        \itc
          \Bigl(
            \card T < k
              \implies
                S = \bigunion \{\,
                                U \in \sh
                              \mid
                                T \sseq U \sseq S
                              \,\}
          \Bigr) \\
    &\qquad\qquad
                \land S \inters V = \emptyset \\
    &\iff
  S \in \rhoTSDk(\sh) \setdiff \rel\bigl(V, \rhoTSDk(\sh)\bigr). \mathproofbox
\end{align*}
\end{proof*}

\begin{lem}
\label{lem:rho-inclusion-implies-rel-star-inclusion}
For each $\sh_1, \sh_2 \in \SH$, each $V \in \wp(\VI)$
and each $k \in \Nset$ with $1 < k \leq n$,
\[
  \rhoTSDk(\sh_1) \sseq \rhoTSDk(\sh_2)
    \implies
      \rel(V, \sh_1)^\star \sseq \rel(V,\sh_2)^\star.
\]
\end{lem}
\begin{proof}
We prove that
\[
  \sh_1 \sseq \rhoTSDk(\sh_2)
    \implies
      \rel(V,\sh_1) \sseq \rel(V,\sh_2)^\star.
\]
The result then follows from Eqs.~(\ref{eq-emi-rho}) and~(\ref{eq-emi-star}).

Suppose $S \in \rel(V, \sh_1)$.
Then, $S \in \sh_1$ and $V \inters S \neq \emptyset$.
By the hypothesis, $S \in \rhoTSDk(\sh_2)$.
Let $x \in V \inters S$.
Then, by Definition \ref{def:rhoSTk}, we have
\begin{align*}
  S &= \bigunion
         \bigl\{\,
           U \in \sh_2
         \bigm|
           \{x\} \sseq U \sseq S
         \,\bigr\} \\
    &= \bigunion
         \bigl\{\,
           U \in \rel(V, \sh_2)
         \bigm|
           \{x\} \sseq U \sseq S
         \,\bigr\}.
\end{align*}
Thus $S \in \rel(V, \sh_2)^\star$.
\end{proof}

\begin{lem}
\label{lem:rho-equiv-implies-amgu-cong}
For each $\sh_1, \sh_2 \in \SH$, each $\sigma \in \Subst$
and each $k \in \Nset$ with $1 \leq k \leq n$,
\[
  \rhoTSDk(\sh_1) = \rhoTSDk(\sh_2)
    \implies
      \rhoTSDk\Bigl(\amgu\bigl(\sh_1, \sigma\bigr)\Bigr)
        = \rhoTSDk\Bigl(\amgu\bigl(\sh_2, \sigma\bigr)\Bigr).
\]
\end{lem}
\pagebreak
\begin{proof}
If $\sigma = \emptyset$, the statement is obvious from the definition
of $\amgu$.
In the other cases, the proof is by induction on the size of $\sigma$.
The inductive step, when $\sigma$ has more than one binding,
is straightforward.
For the base case, when $\sigma = \{ x \mapsto t \}$,
we have to show that
\[
  \sh_1 \sseq \rhoTSDk(\sh_2)
    \implies
      \amgu\bigl(\sh_1, \{x \mapsto t\}\bigr)
        \sseq \rhoTSDk\Bigl(\amgu\bigl(\sh_2, \{x \mapsto t\}\bigr)\Bigr).
\]
The result then follows from Eq.~(\ref{eq-emi-rho}).

Let $v_x \defeq \{x\}$, $v_t \defeq \vars(t)$,
and $v_{xt} \defeq v_x \union v_t$.
Suppose
\[
  S \in \amgu\bigl(\sh_1, \{x \mapsto t\}\bigr).
\]
Then, by definition of $\amgu$,
\[
  S \in \bigl(\sh_1 \setdiff \rel(v_x \union v_t, \sh_1)\bigr)
    \union
      \bin\bigl(\rel(v_x, \sh_1)^\star,
                \rel(v_t, \sh_1)^\star\bigr).
\]
There are two cases:
\begin{enumerate}
\item[1.]
\(
  S \in \sh_1 \setdiff \rel(v_x \union v_t, \sh_1)
\).
Then, by hypothesis, $S \in\rhoTSDk(\sh_2)$.
Hence we have
$S \in \rhoTSDk(\sh_2) \setdiff \rel\bigl(v_x \union v_t,\rhoTSDk(\sh_2)\bigr)$.
Thus, by Lemma~\ref{lem:addit},
\[
  S \in \rhoTSDk\bigl(\sh_2 \setdiff \rel(v_x \union v_t, \sh_2)\bigr).
\]
\item[2.]
\(
  S \in \bin\bigl(\rel(v_x, \sh_1)^\star,
                  \rel(v_t, \sh_1)^\star\bigr)
\).
Then we must have $S = T \union R$ where
$T \in \rel(v_x, \sh_1)^\star$ and
$R \in \rel(v_t, \sh_1)^\star$.
\end{enumerate}

The proof here splits into two branches, 2a and 2b, 
depending on whether $k> 1$ or $k=1$.
\begin{enumerate}
\item[2a.]
We first assume that $k > 1$.
Then, by Lemma~\ref{lem:rho-inclusion-implies-rel-star-inclusion} we have that
$T \in \rel(v_x, \sh_2)^\star$ and
$R \in \rel(v_t, \sh_2)^\star$.
Hence,
\[
  S \in \bin\bigl(\rel(v_x, \sh_2)^\star,
                  \rel(v_t, \sh_2)^\star\bigr).
\]
\end{enumerate}

Combining case~1 and case~2a we obtain
\[
  S \in \rhoTSDk\bigl(\sh_2 \setdiff \rel(v_x \union v_t, \sh_2)\bigr)
    \union  \bin\bigl(\rel(v_x, \sh_2)^\star,
                      \rel(v_t, \sh_2)^\star\bigr).
\]
Hence as $\rhoTSDk$ is extensive and monotonic
\[
  S \in \rhoTSDk\Bigl(\bigl(\sh_2 \setdiff \rel(v_x \union v_t, \sh_2)\bigr)
    \union \bin\bigl(\rel(v_x, \sh_2)^\star,
                     \rel(v_t, \sh_2)^\star\bigr)\Bigr),
\]
and hence, when $k>1$,
$S \in \rhoTSDk\Bigl(\amgu\bigl(\sh_2, \{x \mapsto t\}\bigr)\Bigr)$.

\begin{enumerate}
\item[2b.]
Secondly suppose that $k=1$.
In this case, we have,  by Proposition~\ref{pro: k-self-union}:
 \begin{align*}
\rhoTSDone(\sh_2) &= \sh_2^\star\\
\intertext{%
 and that 
}
\rhoTSDone\Bigl(\amgu\bigl(\sh_2, \{x \mapsto t\}\bigr)\Bigr) 
     &= \amgu\bigl(\sh_2, \{x \mapsto t\}\bigr)^\star.
\end{align*}
Thus, by the hypothesis,
\begin{align*}
  S &\in \bin\bigl(\rel(v_x, \sh_2^\star)^\star,
                  \rel(v_t, \sh_2^\star)^\star\bigr),\\
    &=  \bin\bigl(\rel(v_x, \sh_2^\star),
                  \rel(v_t, \sh_2^\star)\bigr).
\end{align*}
Therefore we can write 
\[
S= T_{\_} \union T_x \union R_{\_} \union R_t
\]
where 
\begin{align*}
T_{\_} \union T_x &\in \rel(v_x,\sh_2^\star),\\
R_{\_} \union R_t &\in \rel(v_t,\sh_2^\star),\\
T_{\_}, R_{\_} &\in \bigl(\sh_2 \setdiff \rel(v_{xt}, \sh_2)\bigr)^\star,\\
T_x &\in \rel(v_x,\sh_2)^\star\setdiff\emptyset,\\
R_t &\in \rel(v_t,\sh_2)^\star\setdiff\emptyset.
\end{align*}
Thus 
\begin{align*}
S &\in \Bigl(\bigl(\sh_2 \setdiff \rel(v_{xt}, \sh_2)\bigr) \union \bin\bigl(\rel(v_x, \sh_2)^\star,
                      \rel(v_t, \sh_2)^\star\bigr)\Bigr)^\star\\
   &= \amgu\bigl(\sh_2, \{x \mapsto t\}\bigr)^\star.
\end{align*}
\end{enumerate}

Combining case~1 and case~2b for $k=1$, 
the result follows immediately
by the monotonicity and extensivity of $(\cdot)^\star$.
\end{proof}

\begin{lem}
\label{lem:rho-equiv-implies-rho-rho-equiv}
For each $\sh_1, \sh_2 \in \SH$,
\[
  \rhoTSDk(\sh_1\union \sh_2)
    =
      \rhoTSDk\bigl(\rhoTSDk(\sh_1) \union \rhoTSDk(\sh_2)\bigr).
\]
\end{lem}
\begin{proof}
This is a classical property of upper closure operators \cite{GHK+80}.
\end{proof}

\begin{lem}
\label{lem:rho-equiv-implies-proj-cong}
For each $\sh_1, \sh_2 \in \SH$ and each $V \sseq \VI$,
\[
  \rhoTSDk(\sh_1) = \rhoTSDk(\sh_2)
   \implies
    \rhoTSDk\bigl(\proj(\sh_1, V)\bigr) = \rhoTSDk\bigl(\proj(\sh_2, V)\bigr).
\]
\end{lem}
\begin{proof}
We show that
\begin{align*}
  \sh_1 \sseq \rhoTSDk(\sh_2)
    \implies
      \proj(\sh_1, V)
        \sseq \rhoTSDk\bigl(\proj(\sh_2, V)\bigr).
\end{align*}
The result then follows from Eq.~(\ref{eq-emi-rho}).

Suppose $\sh_1 \sseq \rhoTSDk(\sh_2)$ and $S \in\proj(\sh_1, V)$.
Then, as $\proj$ is monotonic,
we have $S \in \proj\bigl(\rhoTSDk(\sh_2), V\bigr)$.
We distinguish two cases.
\begin{enumerate}
\item
There exists $x \in V$ such that $S = \{ x \}$.
Then $S \in \proj(\sh_2, V)$ and hence, by Definition~\ref{def:rhoSTk},
$S \in \rhoTSDk\bigl(\proj(\sh_2, V)\bigr)$.
\item
Otherwise, by definition of $\proj$ and Definition~\ref{def:rhoSTk},
there exists $S' \in \rhoTSDk(\sh_2)$ such that $S = S' \inters V$ and
\begin{align*}
  \forall T \sseq S'
    &\itc
      \Bigl(
        \card T < k
          \implies
            S = \bigunion \{\,
                            U \in \sh_2
                          \mid
                            T \sseq U \sseq S'
                          \,\}
                 \inters V
      \Bigr). \\
\intertext{%
Hence
}
  \forall T \sseq S
    &\itc
      \Bigl(
        \card T < k
          \implies
            S = \bigunion \bigl\{\,
                            U \in \proj(\sh_2, V)
                          \bigm|
                            T \sseq U \sseq S
                          \,\bigr\}
      \Bigr),
\end{align*}
and thus $S \in \rhoTSDk\bigl(\proj(\sh_2, V)\bigr)$.
\end{enumerate}
\end{proof}

\begin{proof}[Proof of Theorem~\ref{thm: no-loss-ops}.]
Statements 1, 2 and 3 follow from Lemmas
\ref{lem:rho-equiv-implies-amgu-cong},
\ref{lem:rho-equiv-implies-rho-rho-equiv} and
\ref{lem:rho-equiv-implies-proj-cong}, respectively.
\end{proof}

The following lemma is also proved
in~\cite{BagnaraHZ97b,BagnaraHZ01TCS}
but we include it here for completeness.
\begin{lem}
\label{lem:18}
Let $\sigma \defeq \{ x_1 \mapsto t_1, \ldots, x_n \mapsto t_n \}$,
where, for each $i = 1$, \dots,~$n$, $t_i$ is a ground term.
Then, for all $\sh \in \SH$ we have
\[
  \amgu(\sh, \sigma)
    = \sh \setdiff \rel\bigl(\{ x_1, \ldots, x_n \}, \sh\bigr).
\]
\end{lem}
\begin{proof*}
If $n = 0$, so that $\sigma = \emptyset$, the statement can be easily
verified after having observed that $\rel(\emptyset, \sh) = \emptyset$.
Otherwise, if $n > 0$, we proceed by induction on $n$.
For the base case, let $n = 1$.
Then
\begin{align*}
  \amgu(\sh, x_1 \mapsto t_1)
    &=
      \sh \setdiff \rel\bigl(\{x_1\}, \sh\bigr) \union
        \bin\Bigl(
              \rel\bigl(\{x_1\}, \sh\bigr)^\star,
              \rel\bigl(\emptyset, \sh\bigr)^\star
            \Bigr) \\
    &=
      \sh \setdiff \rel\bigl(\{x_1\}, \sh\bigr).
\end{align*}
For the inductive step, let $n > 1$ and let
\[
  \sigma' \defeq  \{ x_1 \mapsto t_1, \ldots, x_{n-1} \mapsto t_{n-1} \}.
\]
By definition of $\amgu$ we have
\begin{align*}
  \amgu(\sh, \sigma)
    &= \amgu\bigl(\sh, \{ x_n \mapsto t_n \} \union \sigma'\bigr) \\
    &= \amgu\Bigl(\amgu\bigl(\sh, \{ x_n \mapsto t_n \}\bigr), \sigma'\Bigr) \\
    &= \amgu\Bigl(\sh \setdiff \rel\bigl(\{x_n\}, \sh\bigr), \sigma'\Bigr) \\
    &= \Bigl(\sh \setdiff \rel\bigl(\{x_n\}, \sh\bigr)\Bigr)
         \setdiff
           \rel\Bigl(
                 \{ x_1, \ldots, x_{n-1} \},
                 \sh \setdiff \rel\bigl(\{x_n\}, \sh\bigr)
               \Bigr) \\
    &= \sh
         \setdiff
           \biggl(
             \rel\bigl(\{x_n\}, \sh\bigr)
               \union
                 \rel\Bigl(
                   \{ x_1, \ldots, x_{n-1} \},
                   \sh \setdiff \rel\bigl(\{x_n\}, \sh\bigr)
                 \Bigr)
           \biggl) \\
    &= \sh \setdiff \rel\bigl(\{ x_1, \ldots, x_n \}, \sh\bigr). \mathproofbox
\end{align*}
\end{proof*}

\begin{proof}[Proof of Theorem~\ref{thm: smallest-domain-for-ops}.]
We assume that $S \in \rhoTSDk(\sh_1) \setdiff \rhoTSDk(\sh_2)$.
(If such an $S$ does not exist we simply swap $\sh_1$ and $\sh_2$.)

Let $C$ denote a ground term and let
\[
\sigma \defeq \{\, x \mapsto C \mid x \in \VI \setdiff S \,\}.
\]
Then, by Lemma~\ref{lem:18}, for $i = 1$, $2$,
we define $\amgu(\sh_i,\sigma) \defeq \sh_i^S$
where
\begin{align*}
  \sh_1^S &\defeq \{\, T \sseq S \mid T \in \sh_1 \,\}, \\
  \sh_2^S &\defeq \{\, T \sslt S \mid T \in \sh_2 \,\}.
\end{align*}

Now, if $\card S = j$ and $j \leq k$,
then we have $S \in \sh_1 \setdiff \sh_2$.
Hence $S \in \sh_1^S \setdiff \sh_2^S$ and we can easily observe that
$S \in \rhoTSj(\sh_1^S)$ but $S \notin \rhoTSj(\sh_2^S)$.

On the other hand, if $\card S = j$ and $j > k$,
then by Definition~\ref{def:rhoSTk} there exists
$T$ with $\card T < k$ such that
\begin{align*}
  S &=     \bigunion \{\, U \in \sh_1^S \mid T \sseq U \,\} \\
\intertext{%
but
}
  S &\Ssgt \bigunion \{\, U \in \sh_2^S \mid T \sseq U \,\} \defeq S'.
\end{align*}
Let $x \in S \setdiff S'$.
We have $h \defeq \card \bigl(T \union \{x\}\bigr) \leq k$
and thus we can observe that
$T \union \{x\} \in \rhoTSh(\sh_1^S)$
but
$T \union \{x\} \notin \rhoTSh(\sh_2^S)$.
\end{proof}

\section{The Meet-Irreducible Elements}
\label{sec:meet-irreducible}

In Section~\ref{sec:decomposition},
we will use the method of Fil\'e and Ranzato~\cite{FileR96}
to decompose the dependency domains $\TSDk$.
In preparation for this, in this section, we identify the
meet-irreducible elements for the domains and state some general results.

We have already observed that $\TSk$ and $\TSDn = \SH$ are dual-atomistic.
However, $\TSDk$, for $k<n$, is not dual-atomistic and we need to
identify the meet-irreducible elements.
In fact, the set of dual-atoms for $\TSDk$ is
\[
  \dAtoms(\TSDk)
    =
      \bigl\{\,
        \SG \setdiff \{S\}
      \bigm|
        S \in \SG,
        \card S \leq k
      \,\bigr\}.
\]
Note that $\card\dAtoms(\TSDk) = \sum_{j=1}^k \binom{n}{j}$.
Specializing this for $k=1$ and $k=2$, respectively, we have
\begin{align*}
  \dAtoms(\Def)
    &=
      \bigl\{\,
        \SG \setdiff \{\{x\}\}
      \bigm|
        x \in \VI
      \,\bigr\}, \\
  \dAtoms(\PSD)
    &=
      \bigl\{\,
        \SG \setdiff \{S\}
      \bigm|
        S \in \pairs(\VI)
      \,\bigr\} \union \dAtoms(\Def),
\end{align*}
and we have $\card\dAtoms(\Def) = n$ and
$\card\dAtoms(\PSD) = n(n+1)/2$.
We present as an example of this the dual-atoms for $\Def$
and $\PSD$ when $n=3$.

\begin{exmp}
\label{ex: dual-atoms-def-sp}
Consider Example~\ref{ex: three-vars-datoms}.
Then the 3 dual-atoms for $\Def$ are $s_1$, $s_2$, $s_3$ and the 6
dual-atoms for $\PSD$ are  $s_1$, \dots,~$s_6$.
Note that these are not all the meet-irreducible elements since sets
that do not contain the sharing group $xyz$ such as $\{x\}$
and $\bot = \rhoDef(\bot) = \emptyset$ cannot be obtained by the
meet (which is set intersection) of a set of dual-atoms.
Thus, unlike $\Con$ and $\PS$, neither $\Def$ nor $\PSD$ are dual-atomistic.
\end{exmp}

Consider next the set $M_k$ of the meet-irreducible elements of $\TSDk$
that are neither the top element $\SG$ nor dual-atoms.
$M_k$ has an element for each sharing group $S \in \SG$
such that $\card S > k$
and each tuple  $T \sslt S$ with $\card T = k$.
Such an element is obtained from $\SG$
by removing all the sharing groups $U$ such that $T \sseq U \sseq S$.
Formally, for $1 \leq k \leq n$,
\[
  M_k
    \defeq
      \bigl\{\,
        \SG \setdiff \{\,
                       U \in \SG
                     \mid
                       T \sseq U \sseq S
                     \,\}
      \bigm|
        T,S \in \SG,
        T \sslt S,
        \card T =k
      \,\bigr\}.
\]
Note that, as there are $\binom{n}{k}$ possible choices for $T$
and $2^{n-k}-1$ possible choices for $S$, we have
$\card M_k = \binom{n}{k}(2^{n-k} -1)$ and
$\card \MI(\TSDk) = \sum_{j=0}^{k-1} \binom{n}{j} + \binom{n}{k} 2^{n-k}$.

We now show that we have identified precisely all the
meet-irreducible elements of $\TSDk$.
\begin{thm}
\label{thm: mi-elements}
If $k \in \Nset$ with $1 \leq k \leq n$,
then
\[
  \MI(\TSDk) = \{ \SG \} \union \dAtoms(\TSDk) \union M_k.
\]
\end{thm}
The proof of this theorem is included at the end of this section.
Here, we illustrate the result for the case when $n=3$.
\begin{exmp}
\label{ex:MI-Def-and-PSD}
Consider again Example~\ref{ex: dual-atoms-def-sp}.
First, consider the domain $\Def$.
The meet-irreducible elements which are not dual-atoms,
besides $\SG$, are the following (see Figure~\ref{fig:MI-Def}):
\begin{align*}
q_1 &= \{\phantom{x,{}} y,z,\phantom{xy,{}} xz,yz,xyz\} \sslt s_1, \\
q_2 &= \{\phantom{x,{}} y,z,xy,\phantom{xz,{}} yz,xyz\} \sslt s_1, &\qquad
  r_1 &= \{\phantom{x,{}} y,z,\phantom{xy,{}}\phantom{xz,{}}yz\}
    \sslt q_1 \inters q_2,\\
q_3 &= \{x,\phantom{y,{}} z,\phantom{xy,{}} xz,yz,xyz\} \sslt s_2, \\
q_4 &= \{x,\phantom{y,{}} z,xy,xz,\phantom{yz,{}} xyz\} \sslt s_2, &
  r_2 &= \{x,\phantom{y,{}} z,\phantom{xy,{}} xz \phantom{,yz}\}
    \sslt q_3 \inters q_4,\\
q_5 &= \{x,y,\phantom{z,{}} xy,\phantom{xz,{}} yz,xyz\} \sslt s_3, \\
q_6 &= \{x,y,\phantom{z,{}} xy,xz,\phantom{yz,{}} xyz\} \sslt s_3, &
  r_3 &= \{x,y,\phantom{z,{}} xy\phantom{{},xz} \phantom{{},yz}\}
    \sslt q_5 \inters q_6.
\end{align*}

Next, consider the domain $\PSD$.
The only meet-irreducible elements that are not dual-atoms,
beside $\SG$, are the following (see Figure~\ref{fig:MI-PSD}):
\begin{align*}
m_1 &= \{x,y,z, \phantom{xy,{}}  xz, yz\phantom{,xyz}\} \sslt s_4\\
m_2 &= \{x,y,z, xy, \phantom{xz,{}}  yz\phantom{,xyz}\} \sslt s_5\\
m_3 &= \{x,y,z, xy, xz  \phantom{,yz} \phantom{,xyz} \} \sslt s_6.
\end{align*}
Each of these lack a pair
and none contains the sharing group $xyz$.
\end{exmp}

\MakeShortTnput{\tnput}
\newcommand{\X}[3][0pt]{\Toval[framesep=-1mm,name=#3]{\scriptstyle\begin{matrix} #2 \end{matrix}}\tnput{\lower #1\hbox{$#3$}}}
\begin{figure}
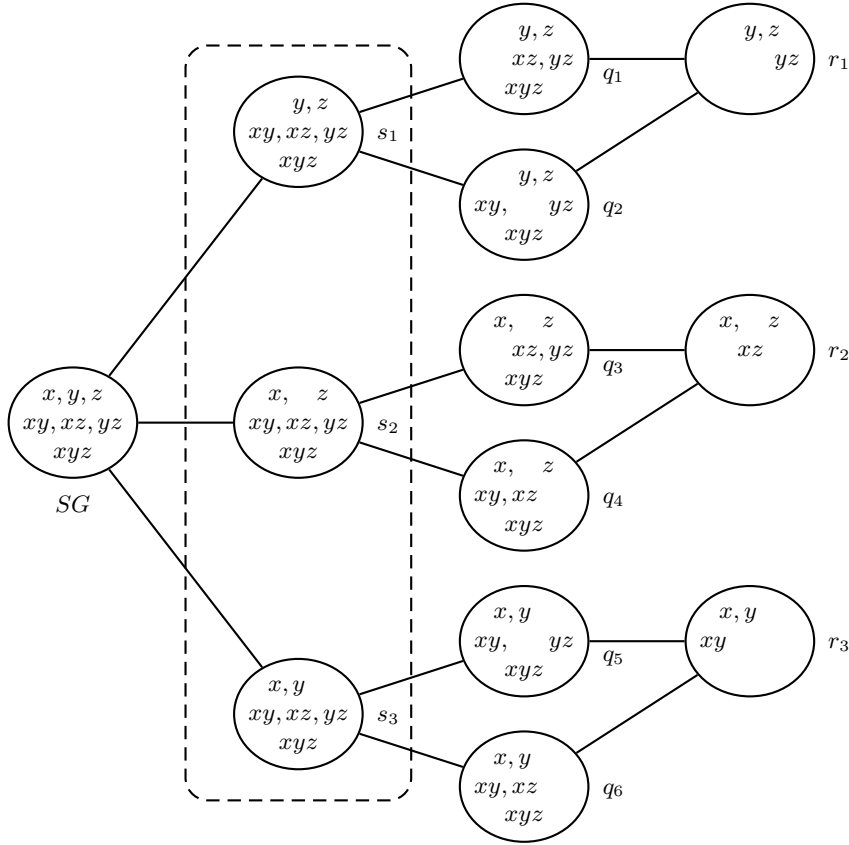

\[
\pstree[treemode=R,treesep=0.44cm,levelsep=3cm]{\X{x,y,z \\ xy,xz,yz \\xyz}{SG}}{%
  \pstree{\X{\phantom{x,{}} y,z \\ xy,xz,yz \\ xyz}{s_1}}{%
    \pstree{\X[5pt]{\phantom{x,{}} y,z \\ \phantom{xy,{}} xz,yz \\ xyz}{q_1}}{%
      \X{\phantom{x,{}} y,z \\ \phantom{xy,{}} \phantom{xz,{}} yz
           \\ \phantom{xyz}}{r_1}
    }
    \X{\phantom{x,{}} y,z \\ xy,\phantom{xz,{}} yz \\ xyz}{q_2}
  }
  \pstree{\X{x,\phantom{y,{}} z \\ xy,xz,yz \\ xyz}{s_2}}{%
    \pstree{\X[5pt]{x,\phantom{y,{}} z \\ \phantom{xy,{}} xz,yz \\ xyz}{q_3}}{%
      \X{x,\phantom{y,{}} z \\ \phantom{xy,{}} xz \phantom{{},yz}
           \\ \phantom{xyz}}{r_2}
    }
    \X{x,\phantom{y,{}} z \\ xy,xz \phantom{{},yz} \\ xyz}{q_4}
  }
  \pstree{\X{x,y\phantom{{},z} \\ xy,xz,yz \\ xyz}{s_3}}{%
    \pstree{\X[5pt]{x,y\phantom{{},z} \\ xy,\phantom{xz,{}} yz \\ xyz}{q_5}}{%
      \X{x,y\phantom{{},z} \\ xy\phantom{{},xz,yz}
           \\ \phantom{xyz}}{r_3}
    }
    \X{x,y\phantom{{},z} \\ xy,xz\phantom{{},yz} \\ xyz}{q_6}
  }
}
\ncline{q_2}{r_1}
\ncline{q_4}{r_2}
\ncline{q_6}{r_3}
\ncbox[linearc=.3,boxsize=1.5,linestyle=dashed,nodesep=.4]{s_1}{s_3}
\]
\caption{The meet-irreducible elements of $\Def$
         for $n = 3$, with dual-atoms emphasized.}
\label{fig:MI-Def}
\end{figure}

\begin{figure}
\[
\pstree[treemode=R,treesep=0.44cm,levelsep=3cm]{\X{x,y,z \\ xy,xz,yz \\xyz}{SG}}{%
  \X{\phantom{x,{}} y,z \\ xy,xz,yz \\ xyz}{s_1}
  \X{x,\phantom{y,{}} z \\ xy,xz,yz \\ xyz}{s_2}
  \X{x,y\phantom{{},z} \\ xy,xz,yz \\ xyz}{s_3}
  \pstree{\X[5pt]{x,y,z \\ \phantom{xy,{}} xz,yz \\ xyz}{s_4}}{%
    \X{x,y,z \\ \phantom{xy,{}} xz,yz \\ \phantom{xyz}}{m_1}
  }
  \pstree{\X[5pt]{x,y,z \\ xy,\phantom{xz,{}} yz \\ xyz}{s_5}}{%
    \X{x,y,z \\ xy,\phantom{xz,{}} yz \\ \phantom{xyz}}{m_2}
  }
  \pstree{\X[5pt]{x,y,z \\ xy,xz\phantom{{},yz} \\ xyz}{s_6}}{%
    \X{x,y,z \\ xy,xz \phantom{{},yz} \\ \phantom{xyz}}{m_3}
  }
}
\ncbox[linearc=.3,boxsize=1.5,linestyle=dashed,nodesep=.4]{s_1}{s_6}
\]
\caption{The meet-irreducible elements of $\PSD$
         for $n = 3$, with dual-atoms emphasized.}
\label{fig:MI-PSD}
\end{figure}

Looking at Examples~\ref{ex: Con-PS-datoms}
and~\ref{ex:MI-Def-and-PSD}, it can be seen that
all the dual-atoms of the domains $\Con$ and $\PS$
are meet-irreducible elements
of the domains $\Def$ and $\PSD$, respectively.
Indeed, the following general result shows that
the dual-atoms of the domain $\TSk$
are meet-irreducible elements for the domain $\TSDk$.
\begin{cor}
\label{cor:mi-TSDk-and-dAtoms-TSk}
Let $k \in \Nset$ with $1 \leq k \leq n$.
Then
\[
  \dAtoms(\TSk) = \bigl\{\, \sh \in \MI(\TSDk) \bigm| \VI \notin \sh \,\bigr\}.
\]
\end{cor}

For the decomposition, we need to identify which
meet-irreducible elements of $\TSDk$ are in $\TSj$.
Using Corollaries~\ref{cor:TSDk-vs-TSj}
and~\ref{cor:mi-TSDk-and-dAtoms-TSk}
we have the following result.
\begin{cor}
\label{cor:MI-TSDk-inters-TSj}
If $j, k \in \Nset$ with $1 \leq j < k \leq n$,
then
\(
  \MI(\TSDk) \inters \TSj = \{ \SG \}.
\)
\end{cor}
By combining Proposition~\ref{prop:TSDk-chain}
with Theorem~\ref{thm: mi-elements}
we can identify the meet-irreducible elements of $\TSDk$ that are in
$\TSDj$, where $j < k$. 
\begin{cor}
\label{cor:MI-TSDk-inters-TSDj}
If
$j, k \in \Nset$ with
$1 \leq j < k \leq n$,
then \[\MI(\TSDk) \inters \TSDj = \dAtoms(\TSDj).\]
\end{cor}

\subsection{Proof of Theorem~\ref{thm: mi-elements}}

\begin{proof}[Proof of Theorem~\ref{thm: mi-elements}.]
We prove the two inclusions separately.
\begin{enumerate}
\item[1.]
$\MI(\TSDk) \Sseq \{\SG\} \union \dAtoms(\TSDk) \union M_k$.

Let $m$ be in the right-hand side. If $m \in \{\SG\} \union \dAtoms(\TSDk)$
there is nothing to prove.
Therefore we assume $m \in M_k$.
We need to prove that if $\sh_1, \sh_2 \in \TSDk$ and
\[
  m = \sh_1 \meet \sh_2 \defeq \sh_1 \inters \sh_2
\]
then $m = \sh_1$ or $m = \sh_2$.
Obviously, we have $m \sseq \sh_1$ and $m \sseq \sh_2$.
Moreover, by definition of $M_k$,
there exist $T, S \in \SG$ where $\card T=k$ and $T \sslt S$ such that
\[
  m = \SG \setdiff \bigl\{\, U \in \SG \bigm| T \sseq U \sseq S \,\bigr\}.
\]
Since $S \notin m$, we have $S \notin \sh_1$ or $S \notin \sh_2$.
Let us consider the first case (the other is symmetric).
Then, applying the definition of $\TSDk$,
there is a $T' \sslt S$ with $\card T' < k$ such that
\[
  \bigunion \{\, U'\in \sh_1 \mid T' \sseq U' \sseq S \,\} \neq S.
\]
Since $\card T' < \card T$, there exists $x$ such that
$x \in T \setdiff T'$. Thus $T' \sslt S\setdiff\{x\}$ and
$S \setdiff \{x\}\in m$. Hence, as $m \sseq \sh_1$,
we have $S \setdiff \{x\}\in \sh_1$.
Consider an arbitrary $U \in \SG$ where $T\sseq U \sseq S$. 
Then $x\in U$. Thus,
since $S = \bigl(S\setdiff\{x\}\bigr) \union U$ and $S\notin \sh_1$,
 $U\notin \sh_1$. 
Thus, as this is true for all such $U$, $\sh_1 \sseq m$.

\item[2.]
$\MI(\TSDk) \sseq \{\SG\} \union \dAtoms(\TSDk) \union M_k$.

Let $\sh \in \TSDk$.
We need to show that $\sh$ is the meet of elements in the right-hand side.
If $\sh = \SG$ then there is nothing to prove.
Suppose $\sh \neq \SG$.
For each $S \in \SG$ such that
$S \notin \sh$, we will show there is an element $m_S$ in the right-hand side
such that $S \notin m_S$ and $\sh \sseq m_S$.
Then $\sh = \biginters \{\, m_S \mid S \notin \sh \,\}$.

There are two cases.
\begin{enumerate}
\item[2a.]
$\card S \leq k$;
Let $m_S = \SG \setdiff \{S\}$.
Then $m_S \in \dAtoms(\TSDk)$ and $\sh \sseq m_S$.

\item[2b.]
$\card S > k$;
in this case, applying the definition of $\TSDk$,
there must exist a set $T' \sslt S$ with $\card T' < k$
such that
\[
   \bigunion \{\, U'\in \sh \mid T' \sslt U' \sseq S \,\} \sslt S.
\]
However, since $T' \sslt S$, we have
$S =
  \bigunion \bigl\{\, T'\union\{x\} \bigm| x \in S\setdiff T' \,\bigr\}$.
Thus, for some $x\in S\setdiff T'$, if
$U$ is such that $T'\union\{x\} \sseq U \sseq S$ then $U \notin \sh$.
Choose $T \in \SG$ so that $T'\union\{x\} \sseq T$ and $\card T = k$ and
let $m_S = \SG\setdiff \{\, U \in \SG \mid T \sseq U \sseq S \,\}$.
Then $m_S \in M_k$, $S\notin m_S$, and $\sh\sseq m_S $.
\end{enumerate}
\end{enumerate}
\end{proof}

\section{The Decomposition of the Domains}
\label{sec:decomposition}

\subsection{Removing the Tuple-Sharing Domains}

We first consider the decomposition of $\TSDk$ with respect to $\TSj$.
It follows from Theorem~\ref{thm:FileR96}
and Corollaries~\ref{cor:TSDk-vs-TSj} and \ref{cor:MI-TSDk-inters-TSj}
that, for $1\leq j < k \leq n$, we have
\begin{align}
   \TSDk \sim \TSj
                 &= \Moore\bigl( \MI(\TSDk)\setdiff\rhoTSj(\TSDk)\bigr)\notag\\
                 &= \Moore\bigl( \MI(\TSDk)\setdiff\TSj\bigr)\notag\\
                 &= \TSDk.
                      \label{eq: TSDk-TSj}
\end{align}
Since $\SH = \TSDn$, we have,
using Eq.~(\ref{eq: TSDk-TSj}) and setting $k=n$,
that, if $j<n$,
\begin{equation}
 \SH \sim \TSj = \SH.
\label{eq: sh-TSj}
\end{equation}
Thus, in general,
$\TSj$ is too abstract to be removed from $\SH$ by means of
complementation.
(Note that here it is required $j < n$, because we have
$\SH \sim \TSn \neq \SH$.)
In particular, letting $j=1$, $2$ (assuming $n>2$) in Eq.~(\ref{eq: sh-TSj}),
we have
\begin{equation}
  \SH \sim \PS = \SH \sim \Con = \SH,
  \label{eq: sh-con}
\end{equation}
showing that $\Con$
and $\PS$ are too abstract to be
removed from $\SH$ by means of complementation.
Also, by Eq.~(\ref{eq: TSDk-TSj}), letting $j=1$ and $k=2$ it follows
that the complement of $\Con$ in $\PSD$ is $\PSD$.

Now consider decomposing $\TSDk$ using $\TSk$.
It follows from Theorem~\ref{thm:FileR96}, Proposition \ref{prop:TSDk-vs-TSk}
and Corollary~\ref{cor:mi-TSDk-and-dAtoms-TSk}
that, for $1 \leq k \leq n$,  we have
\begin{align}
   \TSDk \sim \TSk
                 &= \Moore\bigl( \MI(\TSDk)\setdiff\rhoTSk(\TSDk)\bigr)\notag\\
                 &= \Moore\bigl( \MI(\TSDk)\setdiff\TSk\bigr)\notag\\
                 &= \{\, \sh\in\TSDk \, \mid \VI \, \in \sh \,\}.
                      \label{eq: TSDk-TSk}
\end{align}
Thus we have
\begin{equation}
         \TSDk \sim (\TSDk \sim \TSk) = \TSk.
            \label{eq: TSDk-TSDk}
\end{equation}
We have therefore extracted \emph{all} the domain $\TSk$
from $\TSDk$.
So by letting $k=1$, $2$ in Eq.~(\ref{eq: TSDk-TSk}),
we have found the complements of $\Con$ in
$\Def$ and $\PS$ in $\PSD$:
\begin{align*}
  \Def \sim \Con &= \{\,\sh \in \Def \mid \VI \in \sh \,\}, \\
  \PSD \sim \PS &= \{\,\sh \in \PSD \mid \VI \in \sh \,\}.
\end{align*}
Thus if we denote the domains induced by these complements as
$\Defoplus$ and $\PSDoplus$, respectively, we have the following result.
\begin{thm}
\begin{alignat*}{2}
         \Def \sim \Con &= \Defoplus,
         &\qquad\Def \sim \Defoplus &= \Con,\\
          \PSD \sim \PS &= \PSDoplus,
         &\qquad\PSD \sim \PSDoplus &= \PS.
\end{alignat*}
Moreover, $\Con$ and $\Defoplus$ form a minimal decomposition
for $\Def$ and, similarly, $\PS$ and $\PSDoplus$ form a minimal decomposition
for $\PSD$.
\end{thm}

\subsection{Removing the Dependency Domains}

First we note that, by Theorem \ref{thm: mi-elements},
Proposition~\ref{prop:TSDk-chain}, and Corollary~\ref{cor:MI-TSDk-inters-TSDj},
the complement of
$\TSDj$ in $\TSDk$, where $1 \leq j < k \leq n$, is given as follows:
\begin{align}
\TSDk \sim \TSDj
          &= \Moore\bigl(\MI(\TSDk)\setdiff \rhoTSDj(\TSDk)\bigr)\notag\\
          &= \Moore\bigl(\MI(\TSDk)\setdiff \TSDj\bigr)\notag\\
          &= \bigl\{\,
               \sh\in\TSDk
             \bigm|
                \forall S\in \SG \itc \card S \leq j \implies S\in \sh
             \,\bigr\}.
           \label{eq: TSDk-TSDj}
 \end{align}
It therefore follows from Eq.~(\ref{eq: TSDk-TSDj}) and setting $k=n$
that the complement of $\rhoTSDj$ in $\SH$
for $j<n$ is:
\begin{align}
    \SH \sim \TSDj
          &= \bigl\{\,
               \sh\in\SH
             \bigm|
                \forall S\in \SG \itc \card S \leq j \implies S\in \sh
             \,\bigr\}
          \label{eq: SH-TSDj}\\
          &\defeq \SHjplus.\notag
           \end{align}
In particular, in Eq. (\ref{eq: SH-TSDj}) when $j = 1$,
we have the following result for $\Def$,
also proved in~\cite[Lemma 5.4]{FileR96}:
\begin{align*}
  \SH \sim \Def
    &=
      \bigl\{\,
        \sh\in\SH
      \bigm|
        \forall x\in \VI \itc \{x\}\in \sh
      \,\bigr\}\\
    &\defeq
      \SHDefplus.
\intertext{%
Also, in Eq. (\ref{eq: SH-TSDj}) when $j = 2$, 
we have the following result for $\PSD$:
}
    \SH \sim \PSD
          &= \bigl\{\,
              \sh \in \SH
            \bigm|
              \forall S\in \SG \itc \card S \leq 2 \implies S\in \sh
            \,\bigr\}\\
    &\defeq
      \SHPSDplus.
\end{align*}

We next construct the complement of $\PSD$ with respect
to $\Def$.
By Eq.~(\ref{eq: TSDk-TSDj}),
\begin{equation*}
\begin{split}
  \PSD \sim \Def
    &=
       \bigl\{\,
         \sh\in \PSD
       \bigm|
         \forall x\in \VI \itc \{x\}\in \sh
       \,\bigr\} \\
    &\defeq \PSDplus.
\end{split}
\end{equation*}
Then the complement factor
$\Defminus \defeq \PSD \sim \PSDplus$
is exactly the same thing as
$\SH \sim \SHDefplus$
so that $\PSD$ and $\SH$ behave similarly for $\Def$.

\subsection{Completing the Decomposition}

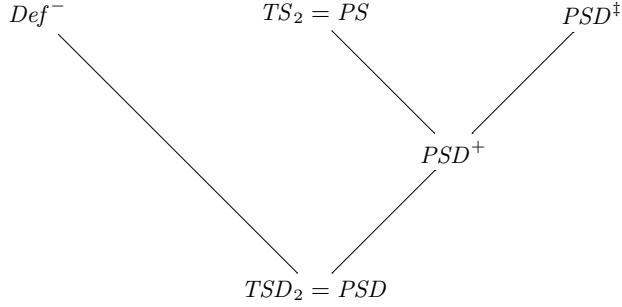
\begin{figure}[t!]
\def\objectstyle{\displaystyle}
\xymatrix@!C=1cm@!R=1cm{
  \Defminus & & \TStwo = \PS & & \PSDplusplus \\
  & & & \PSDplus \ar@{-}[ul] \ar@{-}[ur] \\
  & & \TSDtwo = \PSD \ar@{-}[ulul] \ar@{-}[ur] \\
}
\caption{A non-trivial decomposition of $\PSD$.}
\label{fig:PSD-decomposition}
\end{figure}

Just as for $\SH$, the complement of $\SHDefplus$ using $\PS$
(or, more generally, $\TSj$ where $1 < j < n$) is $\SHDefplus$.
By Corollary~\ref{cor:mi-TSDk-and-dAtoms-TSk}
and Theorem~\ref{thm:FileR96},
as $\PS$ is dual-atomistic,
the complement of $\PS$ in $\PSD^+$ is given as follows.
\begin{thm}
\begin{align*}
  \PSDplusplus
    &\defeq
      \PSDplus \sim \PS \\
    &=
      \bigl\{\,
        \sh \in \PSD
      \bigm|
        \VI \in \sh, \forall x \in \VI: \{x\} \in \sh
      \,\bigr\}, \\
  \PSDplus \sim \PSDplusplus
    &=
      \PS.
\end{align*}
\end{thm}
So, we have extracted \emph{all} the domain $\PS$
from $\PSDplus$ and we have the following result
(see Figure~\ref{fig:PSD-decomposition}).
\begin{cor}
    $\Defminus$, $\PS$, and $\PSDplusplus$ form a minimal decomposition
    for $\PSD$.
\end{cor}

\section{Discussion}
\label{sec:discussion}

By studying the sharing domain $\SH$ in a more general framework,
we have been able to show that the domain $\PSD$ has a natural place
in a scheme of domains based on $\SH$.
Since the well-known domain $\Def$ for groundness analysis
is an instance of this scheme,
we have been able to highlight the close relationship
between $\Def$ and $\PSD$
and the many properties they share. In particular,
it was somehow unexpected that these domains could both be
obtained as instances of a single parametric construction.
As another contribution, we have generalized and strengthened the results
in~\cite{CortesiFW94,CortesiFW98} and~\cite{BagnaraHZ97b,BagnaraHZ01TCS}
stating that
\begin{itemize}
\item
$\Def$ is the quotient of $\SH$ with respect to
the groundness domain $G \equiv \Con$; and
\item
$\PSD$ is the quotient of $\SH$ with respect to
the reduced product $\Con \ucomeet \PS$ of groundness and pair-sharing.
\end{itemize}

In the view of recent results
on abstract domain completeness~\cite{GiacobazziR97},
these points can be restated by saying that $\Def$ and $\PSD$ are
the \emph{least fully-complete extensions} (lfce's) of $\Con$ and
$\Con \ucomeet \PS$ with respect to $\SH$, respectively.

From a theoretical point of view,
the quotient of an abstract domain 
with respect to a property of interest
and
the least fully-complete extension of this same property
with respect to the given abstract domain
are not equivalent.
While the lfce is defined for any semantics given by means of continuous
operators over complete lattices,
it is known \cite{CortesiFW94,CortesiFW98} that the quotient may not exist.
However, it is also known \cite{GiacobazziSR98b} that
when the quotient exists it is exactly the same as the lfce,
so that the latter has also been called \emph{generalized quotient}.
In particular, for all the domains considered in this paper,
these two approaches to the completeness problem in abstract interpretation
are equivalent.

In~\cite{BagnaraHZ97b,BagnaraHZ01TCS},
we wrote that $\PSD \sim \PS \neq \PSD$.
This paper now clarifies that statement.
We have provided a minimal decomposition for $\PSD$ whose components
include $\Defminus$ and $\PS$.
Moreover, we have shown that $\Def$ and $\PSD$ are \emph{not} dual-atomistic
and we have completely specified their meet-irreducible elements.
Our starting point was the work of Fil\'e and Ranzato.
In \cite{FileR96}, they noted, as we have,
that $\SHDefplus \sim \PS = \SHDefplus$
so that nothing of the domain $\PS$
could be extracted from $\SHDefplus$.
They observed that $\rhoPS$ maps all dual-atoms that contain
the sharing group $\VI$ to the top element $\SG$ and
thus lose all pair-sharing information.
To avoid this, they replaced the classical pair-sharing domain $\PS$
with the domain $\PS'$ where,
for all $\sh \in \SHDefplus$,
\[
  \rhoPSprime(\sh) = \rhoPS(\sh) \setdiff \bigl(\{\VI\} \setdiff \sh\bigr),
\]
and noted that
$\SHDefplus \sim \PS' = \{\, \sh\in \SHDefplus \mid \VI \in \sh \,\}$.
To understand the nature of this new domain $\PS'$, we first observe that,
\[
  \PS' = \PS \ucomeet \TSn.
\]
This is because
$\TSn = \MI(\TSn) = \bigl\{ \SG \setdiff\{\VI\}, \SG \bigr\}$.
In addition,
\[
  \SHDefplus \sim \TSn = \{\, \sh\in \SHDefplus \mid \VI \in \sh \,\},
\]
which is precisely the same as $\SHDefplus \sim \PS'$.
Thus, since $\SHDefplus \sim \PS = \SHDefplus$,
it is not surprising that it is precisely the \emph{added}
component $\TSn$ that is removed when we compute the complement for
$\SHDefplus$ with respect to $\PS'$.

We would like to point out that, in our opinion,
the problems outlined above
are not the consequence of the particular domains considered.
Rather, they are mainly related to the methodology for decomposing a domain.
As shown here, complementation \emph{alone} is
not sufficient to obtain \emph{truly minimal} decompositions of domains.
The reason being that complementation only depends on the domain's data
(that is, the domain elements and the partial order relation modeling
their intrinsic precision), while it is completely independent from
the domain operators that manipulate that data.
In particular, if the concrete domain contains elements that are redundant
with respect to its operators (because the observable behavior of these
elements
is exactly the same in all possible program contexts) then any factorization
of the domain obtained by complementation will encode this redundancy.
However, the theoretical solution to this problem is well-known
\cite{CortesiFW94,CortesiFW98,GiacobazziR97,GiacobazziSR98b}
and it is straightforward to improve the methodology so as to obtain
truly minimal decompositions: \emph{first} remove all redundancies
from the domain (this can be done by computing the quotient of the domain
with respect to the observable behavior)
and only \emph{then} decompose it by complementation.
This is precisely what is done here.

We conclude our discussion about complementation with a few remarks.
It is our opinion that, from a theoretical point of view,
complementation is an excellent concept to work with:
by allowing the splitting of complex domains into simpler components,
avoiding redundancies between them,
it really enhances our understanding of the domains themselves.

However, as things stand at present, complementation has never been
exploited from a practical point of view.
This may be because it is easier to implement a single complex domain
than to implement several simpler domains and integrate them together.
Note that complementation requires the implementation of a
full integration between components
(i.e., the reduced product together with its corresponding
best approximations of the concrete semantic operators),
otherwise precision would be lost and
the theoretical results would not apply.

Moreover, complementation appears to have little relevance
when trying to design or evaluate better implementations
of a known abstract domain.
In particular, this reasoning applies to the use of complementation
as a tool for obtaining space saving representations for domains.
As a notable example, the GER representation for $\Pos$ \cite{BagnaraS99}
is a well-known domain decomposition that does enable
significant memory and time savings with no precision loss.
This is not (and could not be) based on complementation.
Observe that the complement of $G$
with respect to $\Pos$ is $\Pos$ itself.
This is because of the isomorphisms
$\Pos \equiv \SH$ \cite{CodishS98} and $G \equiv \Con \defeq \TSone$
so that, by Eq.~(\ref{eq: sh-con}), $\Pos \sim G = \Pos$.
It is not difficult to observe that the same phenomenon happens if one
considers the groundness equivalence component $E$, that is,
$\Pos \sim E = \Pos$.
Intuitively, each element of the domain $E$ defines a partition
of the variable of interest $\VI$ into groundness equivalence classes.
In fact, it can be shown that two variables $x,y \in \VI$
are ground-equivalent in the abstract element $\sh \in \SH \equiv \Pos$
if and only if $\rel\bigl(\{x\}, \sh\bigr) = \rel\bigl(\{y\}, \sh\bigr)$.
In particular, this implies both $\{x\} \notin \sh$ and $\{y\} \notin \sh$.
Thus, it can be easily observed that in all the dual-atoms of $\Pos$
no variable is ground-equivalent to another variable
(because each dual-atom lacks just a \emph{single} sharing group).

A new domain for pair-sharing analysis 
has been defined in~\cite{Scozzari00} as
\[
  \ScozzariShPSh = \PSDplus \ucomeet A,
\]
where the $A$ component is a strict abstraction
of the well-known groundness domain $\Pos$.
It can be seen from the definition that
$\ScozzariShPSh$ is a close relative of $\PSD$.
This new domain is obtained, just as in the case for $\PSD$,
by a construction
that starts from the set-sharing domain $\SH \equiv \mathsf{Sh}$
and aims at deriving the pair-sharing information
encoded by $\PS \equiv \mathsf{PSh}$.
However, instead of applying the generalized quotient operator
used to define $\PSD$,
the domain $\ScozzariShPSh$ is obtained by applying
a new domain-theoretic operator
that is based on the concept of \emph{optimal semantics}
\cite{GiacobazziSR98a}.

When comparing $\ScozzariShPSh$ and $\PSD$,
the key point to note is that $\ScozzariShPSh$ is neither
an abstraction nor a concretization of the starting domain $\SH$.
On the one hand $\ScozzariShPSh$
is strictly more precise for computing pair-sharing,
since it contains formulas of $\Pos$ that are not in the domain $\SH$.
On the other hand $\SH$ and $\PSD$ are strictly more precise
for computing groundness, since $\ScozzariShPSh$ does not
contain all of $\Def$: in particular, 
it does not contain any of the elements in $\Con$.

While these differences are correctly stated in~\cite{Scozzari00},
the informal discussion goes further.
For instance, it is argued in~\cite[Section 6.1]{Scozzari00} that
\begin{quotation}
``in [\cite{BagnaraHZ01TCS}] the domain $\PSD$ is compared
to its proper abstractions only,
which is a rather restrictive hypothesis\ldots''
\end{quotation}
This hypothesis is not one that was made in \cite{BagnaraHZ01TCS}
but is a distinctive feature of the generalized quotient approach itself.
Moreover, such an observation is not really appropriate
because, when devising the $\PSD$ domain,
the goal was to \emph{simplify} the starting domain $\SH$
without losing precision on the observable $\PS$.
This is the objective of the generalized quotient operator
and, in such a context, the ``rather restrictive hypothesis''
is not restrictive at all.

The choice of the generalized quotient can also provide several
advantages that have been fully exploited in~\cite{BagnaraHZ01TCS}.
Since an implementation for $\SH$ was available,
the application of this operator resulted in
an \emph{executable} specification of the simpler domain $\PSD$.
By just optimizing this executable specification it was possible
to arrive at a much more efficient implementation:
exponential time and space savings have been achieved
by removing the redundant sharing groups from the computed elements and
by replacing the star-union operator with the 2-self-union operator.
Moreover, the executable specification inherited
all the correctness results readily available for that
implementation of $\SH$, so that the only new result
that had to be proved was the correctness of the optimizations.

These advantages do not hold for the domain $\ScozzariShPSh$.
In fact, the definition of a feasible representation
for its elements and, a fortiori, the definition of
an executable specification of the corresponding abstract operators
seem to be open issues.\footnote{In \cite{Scozzari00},
the only representation given for the elements of $\ScozzariShPSh$
is constituted by infinite sets of substitutions.}
Most importantly, the required correctness results
cannot be inherited from those of $\SH$.
All the above reasons indicate that the generalized quotient
was a sensible choice when looking for a domain simpler than $\SH$
while preserving precision on $\PS$.

Things are different if the goal is
\emph{to improve the precision of a given analysis}
with respect to the observable,
as was the case in~\cite{Scozzari00}.
In this context the generalized quotient would be
the wrong choice, since by definition it cannot help,
whereas the operator defined in~\cite{Scozzari00} could be useful.

\section{Conclusion}
\label{sec:conclusion}

We have addressed the problem of deriving a non-trivial decomposition
for abstract domains tracking groundness and sharing information
for logic languages by means of complementation.
To this end, we have defined a general schema of domains approximating
the set-sharing domain of Jacobs and Langen and we have generalized and
strengthened known completeness and minimality results.
From a methodological point of view, our investigation has shown that,
in order to obtain truly minimal decompositions of abstract interpretation
domains, complementation should be applied to a reference domain already
enjoying a minimality result with respect to the observable property.

\section*{Acknowledgment}

We recognize the hard work required to review technical papers
such as this one and would like to express our real gratitude
to the Journal referees for their critical reading and constructive
suggestions for preparing this improved version.


\hyphenation{ Bruy-noo-ghe Fa-la-schi Her-men-e-gil-do Lu-ba-chev-sky
  Pu-ru-sho-tha-man Roe-ver Ros-en-krantz Ru-dolph }

\end{document}